\documentclass[aps, twocolumn, pre, floatfix]{revtex4}
\usepackage{graphicx,titlesec,amssymb, amsmath, paralist, hyperref, color}
\pdfoutput=1 
\newcommand{\new}[1]{#1}

\def\*#1{\mathbf{#1}}
\begin{document}

\title{Buckling and metastability in membranes with dilation arrays}
\author{Abigail Plummer}
\email{plummer@g.harvard.edu}
\author{David R. Nelson}
\email{drnelson@fas.harvard.edu}
\affiliation{Department of Physics, Harvard University, 17 Oxford Street, Cambridge, Massachusetts 02138, USA}
\date{\today}

\begin{abstract}
We study periodic arrays of impurities that \new{create localized regions of expansion}, embedded in two-dimensional crystalline membranes. These arrays provide a simple elastic model of shape memory. As the size of each dilational impurity increases (or the relative cost of bending to stretching decreases), it becomes energetically favorable for each of the $M$ impurities to buckle up or down into the third dimension, thus allowing for of order $2^M$ metastable surface configurations \new{corresponding to different impurity ``spin" configurations.} With both discrete simulations and the nonlinear continuum theory of elastic plates, we explore the buckling of both isolated dilations and dilation arrays at zero temperature, guided by analogies with Ising antiferromagnets. We conjecture ground states for systems with triangular and square impurity superlattices, and comment briefly on the possible behaviors at finite temperatures.
\end{abstract}

\maketitle

\section{INTRODUCTION}
In this work, we study a periodic array of dilational impurities, placed within a deformable network that approximates a thin elastic sheet. \new{Each impurity locally expands the surface, and, past a buckling threshold, becomes a site of bistability.} The impurities can then relieve in-plane stresses by escaping up or down into the third dimension, \new{with the preferred direction influenced by interactions with nearby impurities}. A complete understanding of such a system would address several important questions in soft matter \new{physics}. 

\new{First, this elastic system can undergo transformations that mimic out-of-plane plastic deformation.} The buckled impurity arrays can be deformed into metastable states indexed by the up and down puckers of $M$ impurities, somewhat analogous to how two-dimensional (2D) crystals in flat space are remodeled by pointlike dislocations pinned at discrete positions determined by a periodic Peierls potential \cite{dislocations}. Beautiful \new{elastic} models exist for visualizing in-plane ``plastic" deformations--- notably, the classic movies and paper by Bragg and Nye using bubbles in a soap solution that are drawn together by capillary forces to form a triangular lattice \cite{bragg}. \new{Developing a simple discrete model for microscopic features that mediate out-of-plane shape changes may suggest productive directions for further understanding plastic deformation into the third dimension.}

Second, a special case of the model studied here was introduced as an elastic model for a shape memory material in the recent work of \citet{oppenheimer}. Inspired by memory effects in crumpled paper, these authors focused on surfaces without preprogrammed target structures that deform in response to sufficiently strong external forces and remain in a deformed state when forces are removed. These shapeable surfaces are of both theoretical and practical interest, with potential applications to soft robotics and deployable structures \cite{seffen, silverberg}. \citet{oppenheimer} suggested the model we are interested in, a lattice of adjacent bistable nodes, to illustrate that simple, ordered systems can display shapeability. Ultimately, they found that more complex lattices with additional sources of frustration resulted in superior shape memory capabilities, and focused their analysis on these more intricate systems. We return here to their simplest model, and its generalizations, in order to work with a system that is easier to treat theoretically.  We extend previous work both by characterizing more precisely the buckling transition of individual impurities \cite{carraro} and also by considering larger arrays with a variety of ordered initial conditions and varying impurity spacing. 

\new{More generally, our model is relevant to efforts to understand how thin sheets can be programmed to assume specific configurations, such as in the context of metamaterials and plant growth \cite{bertoldi, longleaf, klein, holmes}.} A number of interesting models with similarities to the one we investigate here have recently been introduced as methods for shaping thin sheets in three dimensions \cite{halftone, seffen, wang, sussman, food}. \new{In particular, our system has many features in common with origami-inspired systems that allow mechanical properties to be tuned by selectively inverting bistable vertices \cite{silverberg, cones, chenorigami}.}

Finally, the model studied here may be of interest to those examining atomically thin materials \cite{hanakata, ruizgarcia}. Experimentally, large impurity atoms such as silicon and germanium in 2D materials such as graphene have been observed to buckle out of plane in a manner qualitatively similar to our model \cite{hofer, tripathi}. Theoretically, interest in impurity disorder embedded in tethered surfaces (2D generalizations of linear polymer chains) dates back to at least the early 1990's \cite{radzihovsky1991, carraro}. Specifically, \citet{radzihovsky} introduced a model similar to the one studied here, with dilational impurities embedded at random positions in an elastic sheet. However, as shown in this work, \textit{ordered} arrays of impurities highlight new effects that do not appear for the disordered case.

We begin our study by investigating the buckling behavior of a single dilational impurity with simulations (Sec. \ref{single}). We then note that the use of periodic boundaries (with stress relaxation) in a single impurity simulation in fact creates an array of image impurities, identically buckled in a same-side ``ferromagnetic" configuration, and develop a general analytic framework in Fourier space to predict the buckling transition point (Sec. \ref{ferro}). We use this framework together with an analogy with Ising models to explore candidate ground states in more general buckled configurations (Sec. \ref{groundstatesec}). Finally, we introduce and characterize a discrete model with a square host lattice, which lacks the geometric frustration of an underlying triangular mesh, and revisit each of the above topics in this system (Sec. \ref{square}).  We conclude by discussing prospects for future work, including speculations about the behavior of periodic arrays of dilations at finite temperatures (Sec. \ref{discussion}).

\new{We briefly summarize the main results of this report: For the small dilational impurities that we study, buckling occurs at a critical value of a dimensionless parameter $\gamma$ that factors in the cost of bending versus stretching, and the size mismatch between the impurity and the host lattice. Interactions between buckled impurities can be long range close to the buckling transition, and are strongest at intermediate values of $\gamma$. We present evidence that the ground state of a buckled array of impurities is a zigzag pattern on a triangular lattice (Fig. \ref{zigzag}), with geometric frustration playing an important role, and an ``antiferromagnetic'' checkerboard pattern on a square lattice (Fig. \ref{checker}).}

\section{SINGLE IMPURITY}
\label{single}
\new{In this section, we first provide some qualitative understanding of the buckling transition for an isolated dilation by looking at the energy of possible real space deformation fields. We then introduce a discrete model that can incorporate multiple dilations, and discuss boundary effects and correspondence to continuum elasticity theory. Finally, using the discrete model for a single impurity, we observe that buckling is controlled by a dimensionless parameter, as suggested by our real space argument, and find that we expect buckled impurities to interact with one another, especially close to the buckling transition.}

A back-of-the-envelope calculation suggests that a threshold exists at which buckling becomes energetically favorable. \new{This calculation requires many approximations, and thus is only presented to build our intuition for the system, rather than to make a precise prediction for the buckling threshold. We will introduce a quantitative Fourier space calculation for buckling in Sec. \ref{ferro}.}

Consider the in-plane displacement field surrounding a dilational impurity in an isotropic solid initially confined to a 2D plane (a discrete version of a dilation is shown in the inset of Fig. \ref{lattice}) \cite{nelsondefects, landau}, 
\begin{equation}
\label{impurity}
u(\*r)= \frac{\Omega_0 \*r}{2 \pi r^2},
\end{equation}
where $\Omega_0$ measures the extra area added to the surface, discussed in more detail in Sec. \ref{discretetri}. We integrate the in-plane stretching energy from $r=\delta$, a microscopic cutoff of order a lattice constant or surface thickness, to $r=\infty$ to get \cite{landau}
\begin{equation}
E_s=\frac{1}{2} \int d^2r (2 \mu u_{ij}^2+ \lambda u_{kk}^2)=\frac{\mu \Omega_0^2}{2\pi \delta^2}= \frac{Y \Omega_0^2}{4(1+\nu)\pi \delta^2},
\label{ur}
\end{equation}
where $\mu$ and $\lambda$ are Lam\'e coefficients, and $Y$ and $\nu$ are the 2D Young's modulus and Poisson's ratio respectively. \new{Since the surface is flat, this state has zero bending energy.}

\new{On the other hand, we can consider a buckled impurity with negligible stretching energy. We assume for simplicity a Gaussian height profile, $f(r)=H_0 e^{-r^2/2\sigma^2}$. This Gaussian ansatz provides an upper bound on the bending energy (an exponentially decaying height profile gives similar results, as shown in Appendix A). Since we would like this state to have approximately zero stretching energy, the buckling length scale $\sigma$ must be comparable to the inclusion width, which is of order a few lattice constants. With this profile, both bending and stretching energies are negligible except within the inclusion core region.}

The bending energy of this height profile is
\begin{equation}
E_b= \frac{\kappa}{2} \int d^2 r \left( \nabla^2 f \right)^2=\frac{\pi \kappa H_0^2}{\sigma^2}.
\label{realbend}
\end{equation}

\new{If we assume that the buckled state relaxes the stretching due to the impurity completely,} we can relate $\Omega_0$ to $H_0$ through the extra surface area generated by the dilation in the Monge representation.
\begin{align}
\Omega_0&= \int d^2 r \left[ \sqrt{1+ \left( \frac{df}{dr} \right)^2} -1 \right] \nonumber \\ &\approx \pi \int dr r  \left(\frac{df}{dr}\right)^2 = \frac{\pi H_0^2}{2},
\end{align}
which leads to $E_b \approx 2 \Omega_0 \kappa/\sigma^2$. Note that in contrast to the stretching energy in Eq. (\ref{ur}), the bending energy is linear in the extra area $\Omega_0$. If we now ask when the energy of the buckled state is lower than the energy of the flat state, we find that it occurs above a critical value of the dimensionless F{\"o}ppl-von K{\'a}rm{\'a}n number $\gamma$, constructed with the impurity size $\Omega_0$,
\begin{equation}
\gamma_c \equiv \frac{Y \Omega_0}{\kappa} \sim \frac{\delta^2}{\sigma^2},
\label{gammaintro}
\end{equation}
which is of order unity. 

\new{Although this argument simplifies the system considerably (we assume the only two possible states are an inflexible sheet and an approximately inextensible sheet), the finding that the transition is controlled by $\gamma$ will be confirmed by a more detailed analysis presented in Sec. \ref{ferro}. For further discussion of the limitations of comparing zero bending energy and zero stretching energy states to understand buckling transitions, see \citet{efrati}.} 

Note that, unlike classic defect problems, such as the buckling of dislocations and disclinations, $\gamma$ does not diverge with system size \cite{seung, carraro}. We note in passing that dilations can also cause buckling in a one-dimensional semiflexible polymer embedded in two dimensions if the endpoints are held fixed. In this case, the transition is controlled by a dimensionless parameter that depends on system size, and the buckled state is a smooth, global deformation, similar to a classic Euler buckling transition \cite{landau}.

\subsection{Discrete model design: triangular host lattice}
\label{discretetri}
Following Refs. \cite{seung} and \cite{kantor}, we build our model from a 2D triangular lattice. An alternative square lattice model with diagonal bonds will be discussed in Sec. \ref{square}. Neighboring nodes at positions $\*r_i$ and $ \*r_j$ are connected by harmonic springs and neighboring triangular faces are penalized when their normals, $\*n_\alpha$ and $\*n_\beta$, are not aligned. The energy of the system is then the sum of stretching and bending energies,
\begin{equation}
\label{discen}
E= \frac{k}{2} \sum_{\langle i,j \rangle} \left( |\*r_i -\*r_j |- a_{ij}\right)^2+\tilde{\kappa} \sum_{\langle \alpha, \beta \rangle} \left(1-\*n_\alpha \cdot \*n_\beta\right),
\end{equation}
where the first sum is over neighboring nodes, and the second is over neighboring faces. The rest length of the springs connecting nodes $i$ and $j$, $a_{ij}$, is varied to insert dilational impurities. With all $a_{ij}$ equal to $a_0$, this energy approximates the bending and stretching energies used in the previous section in the continuum limit with $k=\frac{\sqrt{3}}{2} Y$, $\tilde{\kappa}= \frac{2}{\sqrt{3}} \kappa$ and a Poisson ratio $\nu=1/3$ \cite{seung}. 

\begin{figure}
\begin{center}
\includegraphics[scale=0.22]{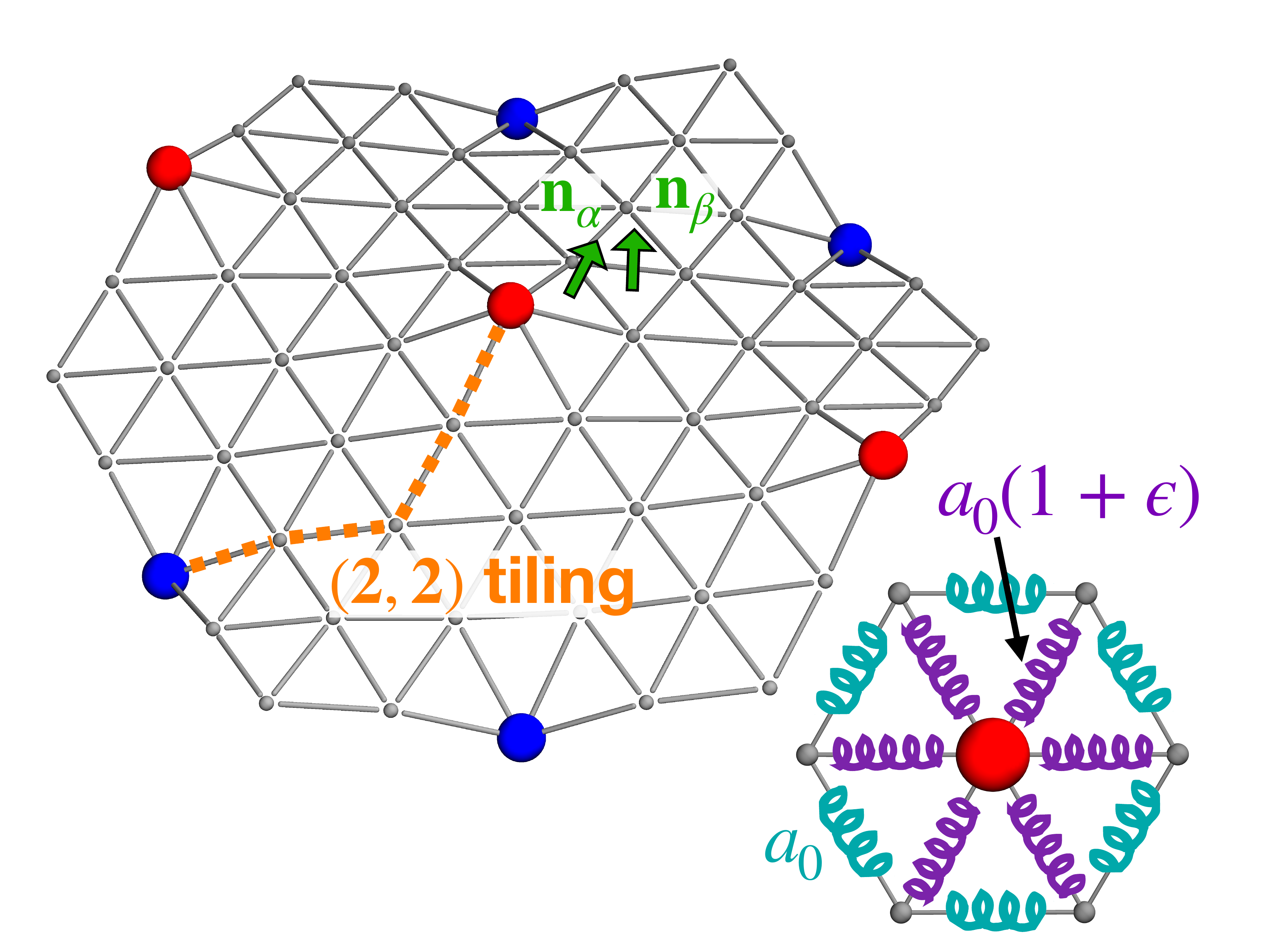}
\end{center}
\caption{{\label{lattice} A schematic defining relevant quantities for our discrete model of dilation arrays. Large nodes with colors mark the impurity sites, here arranged in a $(2,2)$ tiling. Impurities that have buckled up are shown in red, and those that buckled down are in blue. Two normals corresponding to neighboring faces that contribute to the bending energy are shown in green. Inset: A top-down view of a single impurity. Bonds with rest lengths $a_0$ and $a_0(1+\epsilon)$ are highlighted in teal and purple, respectively.
}}
\end{figure}

Next, we insert a dilational impurity at node $i$ by setting $a_{ij}=a_0(1+\epsilon)$ for all neighboring nodes $j$, thus making the rest length of the springs connecting $i$ to the rest of the lattice an amount $a_0 \epsilon$ longer (see inset of Fig. \ref{lattice}).  \new{We restrict our simulations to small positive values of $\epsilon$ in order to minimize anisotropic lattice effects and to work in the same limit as our continuum theory, which will neglect terms of order $\epsilon^2$.}

We describe a periodic array of impurities by a pair of integers $(n,m)$, such that one moves between isolated dilations by taking $n$ steps along one lattice direction, turning counterclockwise by $60^{\circ}$, and taking $m$ more steps, as shown for $(n,m)=(2,2)$ in Fig. \ref{lattice}. The energy minimizations used to explore the metastable configurations in this paper were performed using FIRE \cite{bitzek}, with key results verified with BFGS \cite{fletcher}. All distances are measured in units of $a_0$, the spacing of the background triangular lattice. 

\begin{figure}
\begin{center}
\includegraphics[scale=0.35]{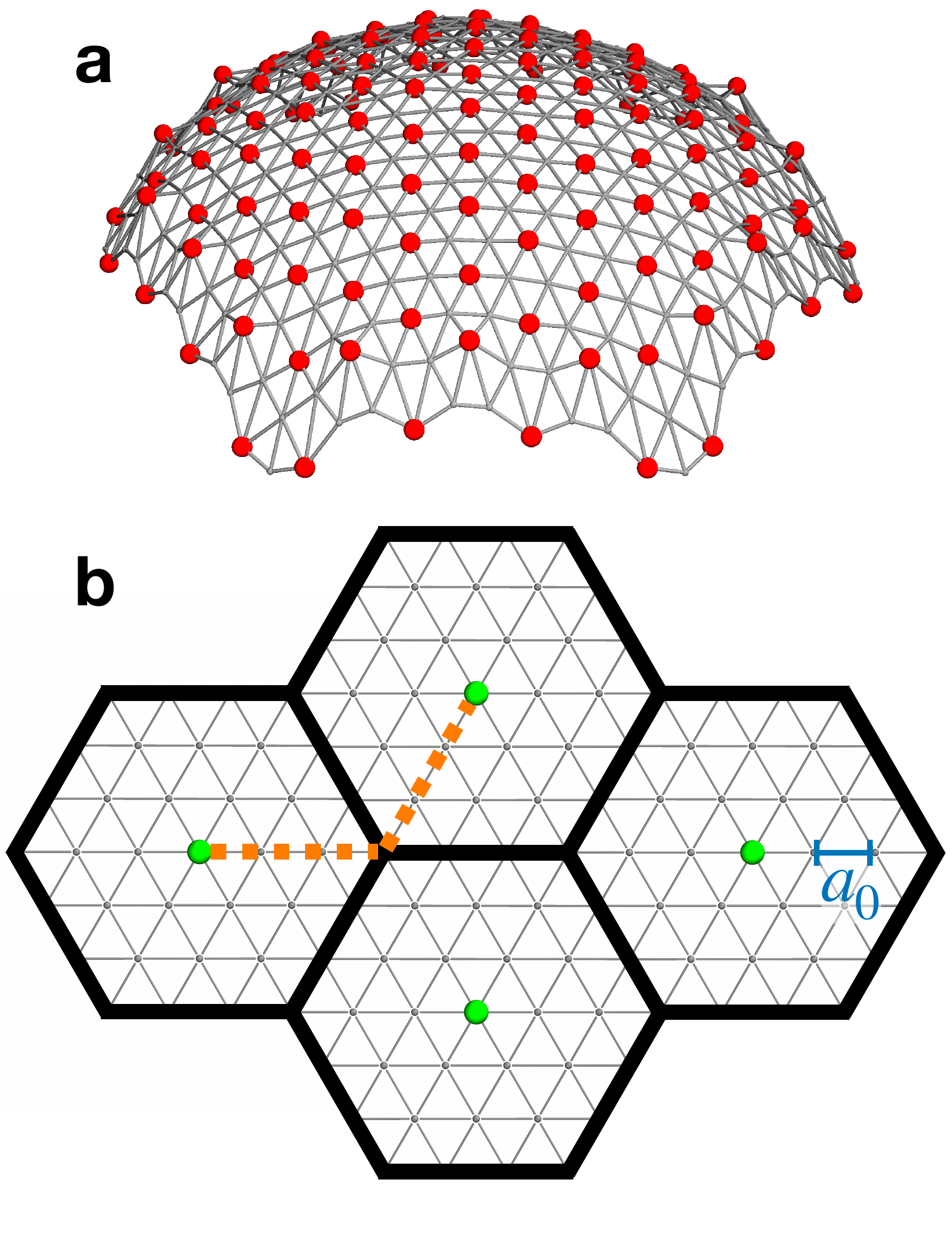}
\end{center}
\caption{{\label{cappbc} (a) A cap-like structure with free boundary conditions found with a simulated annealing protocol, somewhat reminiscent of the chainmail structures of Refs. \cite{klotz, grosberg}. (b) Periodic boundary conditions on a hexagonal simulation cell with a single impurity at the center and radius $R=n a_0$ produce image impurities arranged in an $(n,n)$ array. Here, four identical simulation cells of radius $R=3 a_0$ are shown separated by thick black lines.
}}
\end{figure}

Boundary effects can be strong, especially close to the buckling transition. Impurities can more easily buckle near free edges, and often buckle on the same side of the host surface and then curl beneath it. We observe a dramatic example of this effect if we place the impurities close to one another in a $(1,1)$ array. As shown in Fig. \ref{cappbc}(a), smooth caplike structures are then preferred, even for membranes that would be too stiff to buckle with the relaxed periodic boundaries used in the remainder of this paper. \new{The formation of a spherical cap due to the insertion of identical isotropic dilational impurities is intriguing enough to merit its own investigation, but we limit our discussion to two brief remarks.}

\new{First,} this behavior bears some resemblance to recently observed hemispherical configurations of kinetoplasts, which are ``chainmail" structures composed of rings of DNA \cite{klotz, grosberg}. It would be interesting if the ``maxicircles" of DNA woven into the chainmail structure composed of DNA ``minicircles" could be related to the dilations studied in this paper. \new{Second, we note similarities to three recent papers that study how isotropic defects with $\epsilon<0$ (corresponding to vacancies or small subsitutional impurities) can be used to design surfaces with a given distribution of curvature \cite{sussman, halftone, kupferman}.}

To avoid these \new{(very interesting) boundary effects,} we implement tension-free periodic boundaries on a hexagonal domain. Domain size is quoted in terms of the radius $R$, which is defined as the distance from the hexagon center to its corner when all bond lengths are equal to $a_0$ [Fig. \ref{cappbc}(b)]. We alternately perform local minimizations and global minimizations. Local minimizations move all nodes according to the gradient of Eq. (\ref{discen}) with respect to vertex positions, with the nodes on the top right edge of the periodic hexagon identified with the nodes on the bottom left edge, etc. Global minimizations take place over all possible affine deformations of the background hexagonal simulation cell. Optimization to a local minimum energy configuration is complete when the magnitude of all components of both the $3N$-dimensional local deformation gradient, where $N$ is the number of nodes that can move independently from one another, and the three-dimensional (3D) global affine deformation gradient are below a threshold value.

In Sec. \ref{ferro}, we will use continuum theory to make quantitative predictions about buckling in a discrete flexible membrane with an array of pointlike dilations. In order to verify these predictions with simulations, we must be able to translate between the continuum limit parameters and corresponding microscopic simulation parameters. As mentioned, the mapping of the continuum parameters $Y$ and $\kappa$ to $k$ and $\tilde{\kappa}$, respectively, has been described for this model \cite{seung}. However, we still must find an expression for continuum parameter $\Omega_0$, the extra area added to the surface in flat space due to the dilation [as in Eq. (\ref{impurity})], in terms of the dilational parameter $\epsilon$. We will show that we can relate these parameters by considering a coarse-grained description of the surface's preferred metric, and verify this relation with simulations.

\begin{figure}
\begin{center}
\includegraphics[width=\columnwidth]{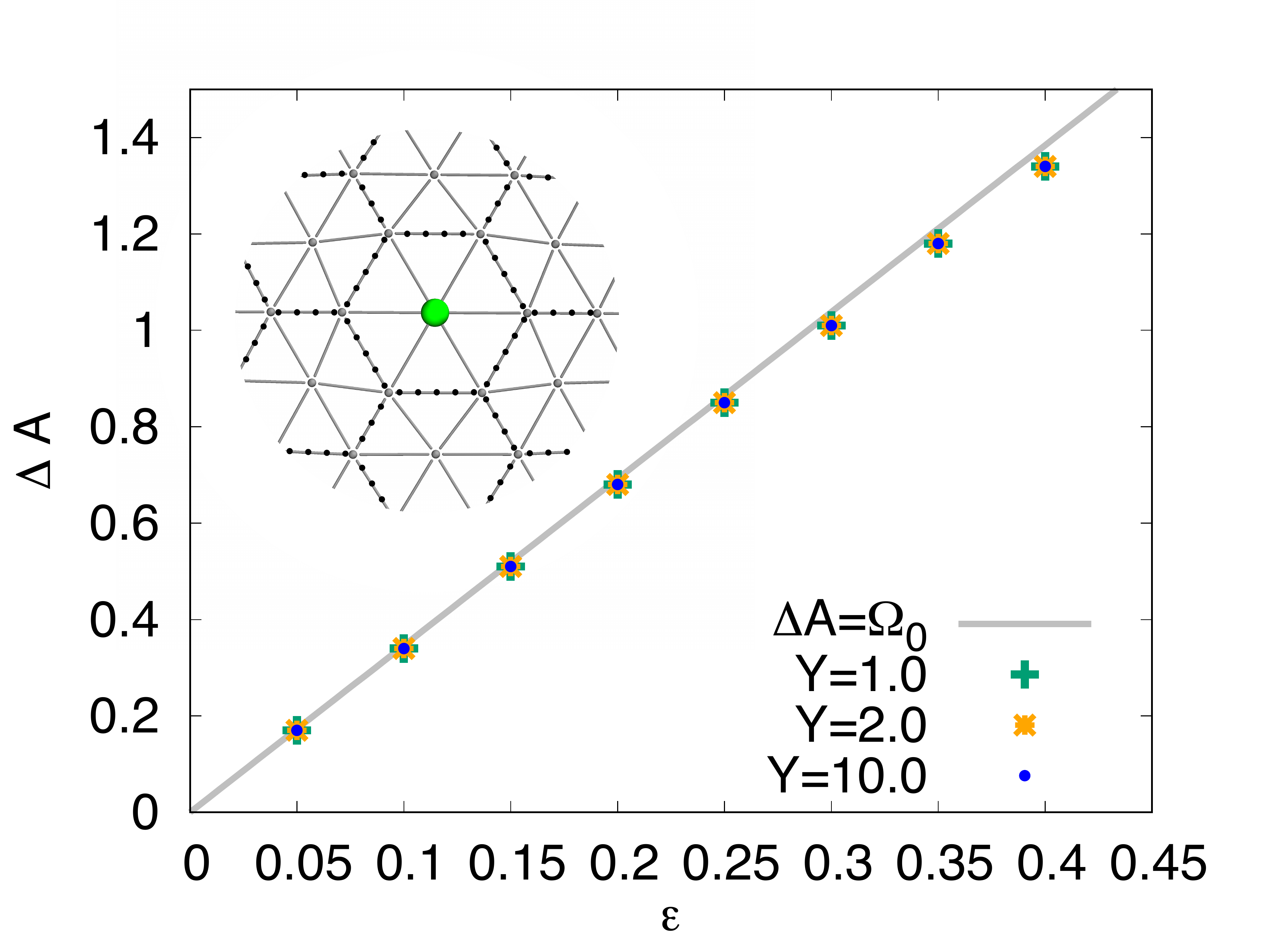}
\end{center}
\caption{{\label{extarea} Comparison between theory and simulation for the macroscopic increase in area due to an impurity vs the parameter $\epsilon$ for various values of the Young's modulus. The gray line shows $\Delta A = 3.46 a_0^2\epsilon$ from Eq. (\ref{omega0}). Data are for a hexagonal periodic cell with radius $R=30a_0$. Inset: Local lattice deformations near an impurity site obtained by energy minimization in the discrete model. Dotted lines show hexagons used to calculate the preferred metric. 
}}
\end{figure}

Consider our discrete lattice with a single impurity at the center of the simulation cell, as in Fig. \ref{cappbc}(b). We divide the mesh into hexagons composed of six triangular faces such that the impurity is at the center of one of the hexagons as in the inset of Fig. \ref{extarea}. Each of these hexagons in isolation has an unambiguously defined preferred surface area that minimizes the stretching associated with the nearest-neighbor springs. The hexagons without the impurity at the center prefer an area $3 a_0^2\sqrt{3}/2$, composed of six equilateral triangles with side length $a_0$, and the hexagon with the impurity prefers a surface area of $3 a_0^2 \sqrt{(1+\epsilon)^2 - 1/4}$, composed of six isosceles triangles with side length $a_0(1+\epsilon)$ and base length $a_0$ (thus forming a 3D prismatic structure). We then define the extra preferred surface area of the impurity hexagon
\begin{align}
\Omega_0&= 3 a_0^2 \sqrt{(1+\epsilon)^2 - 1/4}-3 a_0^2\sqrt{3}/2, \nonumber
\\&\approx 3.464 a_0^2 \epsilon- 0.577 a_0^2 \epsilon^2 +O(\epsilon^3).
\label{omega0}
\end{align}
Note that the extra area $\Omega_0$ scales linearly with the dilation parameter $\epsilon$ for small $\epsilon$. Upon placing the impurity at the origin, we can summarize this description in the continuum limit in terms of a preferred metric for the surface, $\overline{g_{\alpha \beta}}$ \cite{radzihovsky1991}:
\begin{equation}
\overline{g_{\alpha \beta}}= \delta_{\alpha \beta} \left(1+\Omega_0 \delta^2(\*r) \right) .
\label{metric}
\end{equation}
Note that we find the correct excess surface area by integrating $\sqrt{\det g_{\alpha \beta}}$ when the actual surface metric is equal to the preferred metric \new{(a zero stretching energy state)}. If we now confine that lattice to flat space, we can use the preferred metric to determine the strain, leading to the elastic energy (to linear order in $\Omega_0$)\cite{radzihovsky1991},
\begin{equation}
E= \frac{1}{2} \int d^2 r \left[2\mu u_{\alpha \beta}^2 +  \lambda u_{\gamma \gamma}^2 - 2\left(\mu+ \lambda\right) \Omega_0 u_{\gamma \gamma}  \delta^2(\*r) \right].
\label{inplane}
\end{equation}
We can now use this expression to establish a correspondence with continuum \new{(linear)} elasticity, since Eq. (\ref{inplane}) is also the elastic energy for an impurity modeled as a source of stress $\sigma_{ij}^{\text{imp.}}$ in an infinite 2D medium \cite{eshelby, nelsonreentrant}.
\begin{equation}
\sigma_{ij}^{\text{imp.}}= \left(\mu+\lambda\right) \Omega_0 \delta_{ij} \delta^2(\*r).
\end{equation}

This form of $\sigma_{ij}^{\text{imp.}}$ indicates that $\Omega_0$ is the change in area due to the impurity when the surface is embedded in two dimensions (\new{neglecting boundary effects}).
\begin{equation}
\Delta A= \Omega_0= \int d^2 r u_{\gamma \gamma}.
\end{equation}
Therefore, a coarse-grained metric description allows us to write down an expression for $\Omega_0$ in terms of the microscopic model geometry, which we check by measuring the macroscopic expansion \new{of the periodic unit cell} in two dimensions.

Figure \ref{extarea} confirms the approximately linear relationship between the extra area $\Delta A$ and $\epsilon$ with the predicted slope. At large values of $\epsilon$, the data are slightly lower than the theory. \new{Since we neglected terms of order $\Omega_0^2$ in our analysis, some deviation at large $\epsilon$ is not surprising.} This relation is independent of the Young's modulus, as expected from Eq. (\ref{omega0}).

\new{Note that the energy in Eq. (\ref{inplane}) has a term directly coupling the impurity stress to the membrane, in contrast to the energy functional used in our back-of-the-envelope calculation, Eq. (\ref{ur}). We were nevertheless able to estimate the energy of the flat dilation with Eq. (\ref{ur}) because we assumed we knew the form of the displacement field and integrated only over the area outside the impurity core, where the reference metric is simply $\overline{g_{\alpha \beta}} = \delta_{\alpha \beta}$.}

\new{\subsection{Results from the discrete model}}
\new{Now that we have fully specified our discrete model, we survey parameter space for a single dilation embedded in a large patch of a triangular host lattice and present key features of the buckling transition.} As anticipated by the simple argument in the beginning of Sec. \ref{single}, the buckling transition occurs at a critical value of the dilation F{\"o}ppl-von K{\'a}rm{\'a}n number $\gamma$, which we can define in terms of both macroscopic elastic parameters and microscopic simulation parameters as
\begin{equation}
\gamma \equiv \frac{Y \Omega_0}{\kappa} \approx \frac{4 (3.46 a_0^2 \epsilon) k}{3 \tilde{\kappa}}.
\end{equation}

\begin{figure*}
\begin{center}
\includegraphics[width=\textwidth]{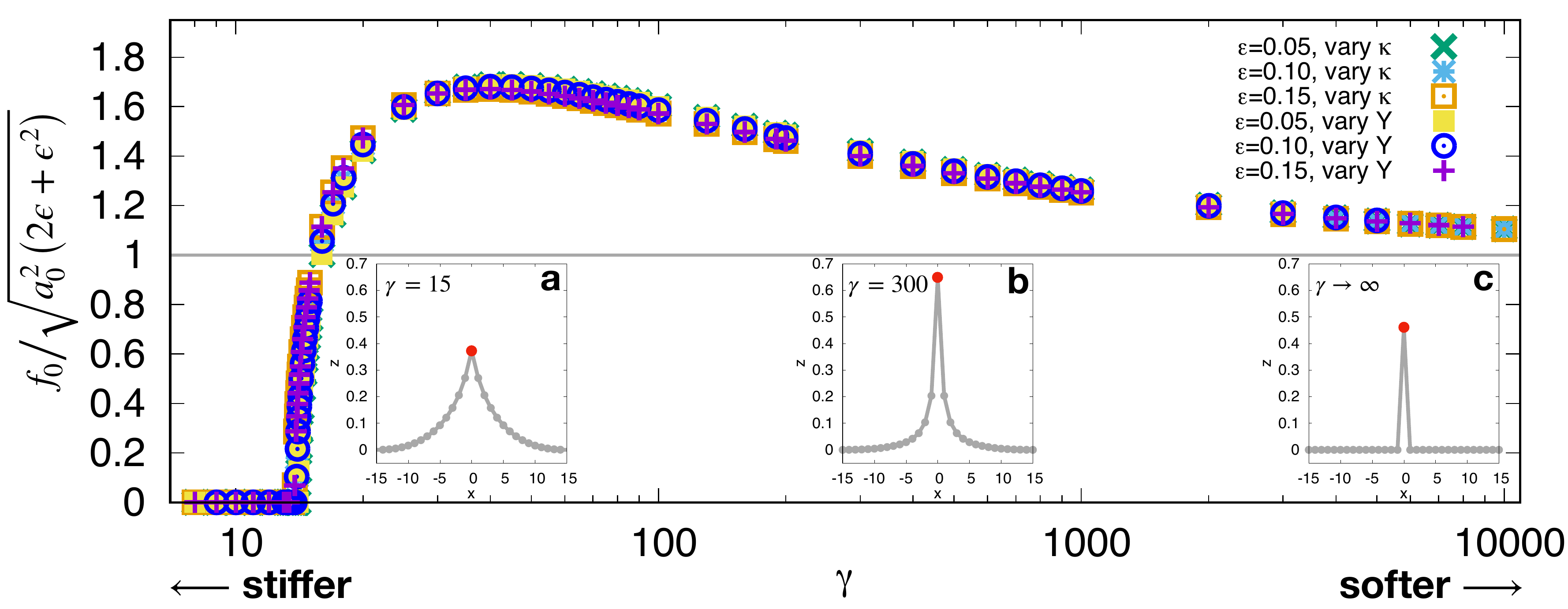}
\end{center}
\caption{{\label{collapse} Height of a single impurity in units of the lattice constant $a_0$, rescaled by the height in the prismatic limit of small $\kappa$ and large $Y$, as a function of the dilation F{\"o}ppl-von K{\'a}rm{\'a}n number $\gamma$. Panels (a)-(c) show the height profile defined in Fig. \ref{height}(a) at specific values of $\gamma$, for $R=15 a_0$. 
}}
\end{figure*}

Important parameter regimes appear in Fig. \ref{collapse}, where we plot the rescaled height $f_0$ of the impurity at the center of our periodic cell as a function of $\gamma$, for fixed system size $R$. \new{The height of the impurity is measured relative to the boundary points, as shown in the panels of Fig. \ref{collapse}.} The rescaling of the height by a function of $\epsilon$ is chosen to match the expected height in the inextensible limit ($\gamma \to \infty$). In this prismatic limit, only the impurity is displaced in the $z$ direction, and all other nodes are undisturbed [Fig. \ref{collapse}(c)]. \new{This is the discrete version of the pure bending state we introduced in our back-of-the-envelope calculation.} The ease with which we can identify the inextensible limit is a nice feature of the discrete model. 

For small $\gamma<\gamma_c \approx 14$, the flat state is energetically favorable. At a critical value of $\gamma_c \approx 14$, the buckled state becomes preferred, with the height of the impurity rising continuously as a function of $\gamma$ from 0 until it reaches a maximum value at $\gamma \approx 40$. In this regime, the height profile of the buckled surface drops off smoothly from the center of the dilation, with vertical displacements slowly going to zero at the boundaries of our periodic hexagonal domain [Fig. \ref{collapse}(a)]. As $\gamma$ continues to increase, the height of the impurity drops [Fig. \ref{collapse}(b)], slowly approaching the prismatic $\gamma \to \infty$ limit [Fig. \ref{collapse}(c)] where the lowest energy configuration is an isolated pyramid centered on the impurity with zero stretching energy.  

\new{We could have anticipated this non-monotonic behavior by considering an impurity in the prismatic state ($\gamma \to \infty$) as we decrease $\gamma$. The finite cost of bending smooths the sharp corners in the height profile, lifting the impurity in the $z$ direction. On the other hand, we can consider an impurity at $\gamma$ just above $\gamma_c$ as we increase $\gamma$. It is plausible that the amplitude of the unstable mode grows continuously as bending becomes less costly (and we will show that this has to be the case in Sec. \ref{ferro}). Since the height of the impurity increases as we move away from either $\gamma_c^+$ or the $\gamma \to \infty$ limit, the height must reach a maximum value at some intermediate $\gamma$.}

Figure \ref{collapse} shows results only for a dilation that has buckled in the positive $z$ direction. We note that states with the same vertical displacements in the negative $z$ direction are energetically equivalent. The buckling direction is selected by the initial condition used in our energy minimization procedure---for each value of $\gamma$, we displace the impurity node at the origin a fixed amount $\Delta z>0$ (all other nodes remain at $z=0$) and then minimize the energy over all nodes as previously described to find the configuration corresponding to a local energy minimum.

\begin{figure}
\begin{center}
\includegraphics[width=\columnwidth]{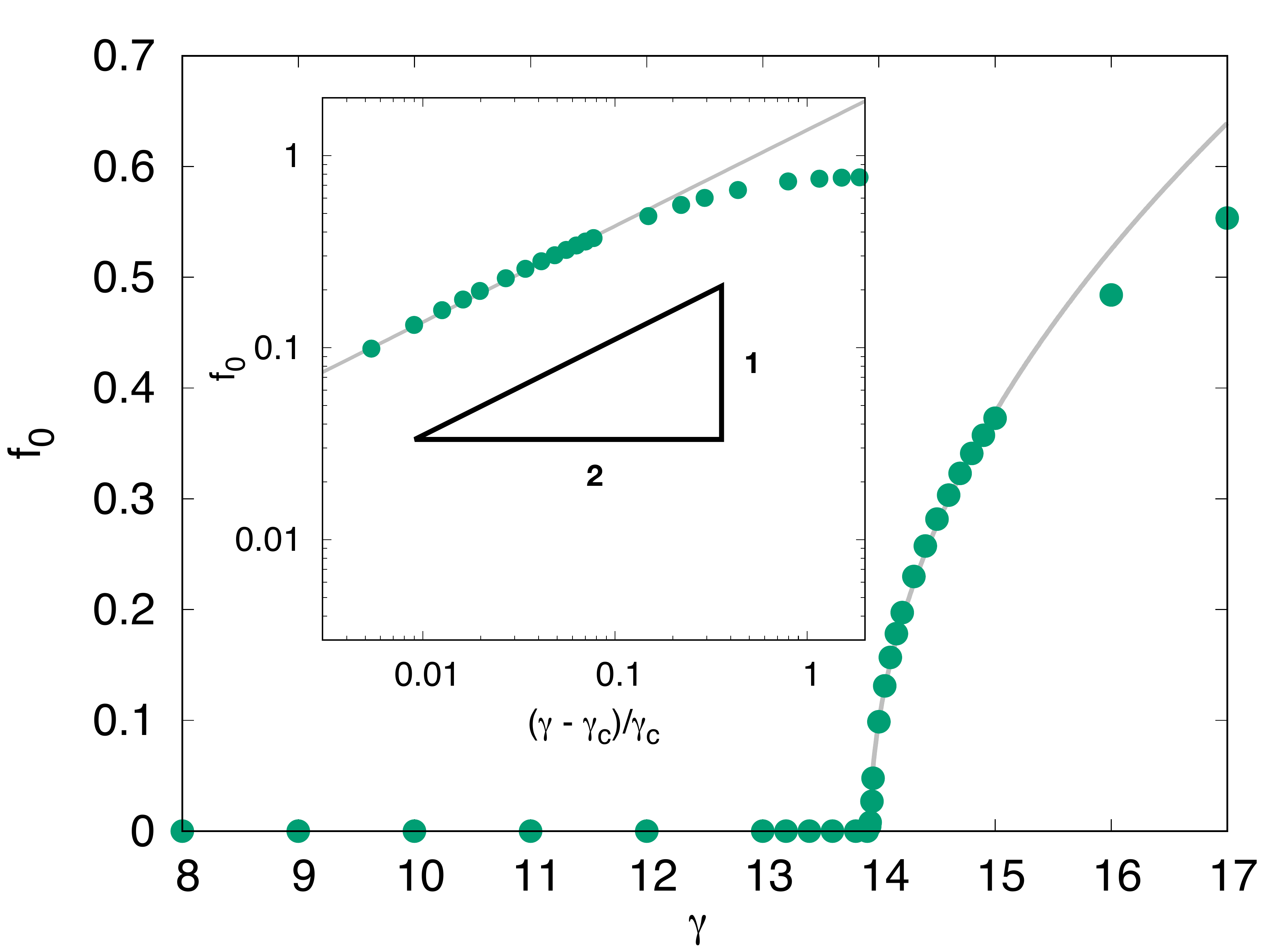}
\end{center}
\caption{{\label{scaling} Close to the transition, the height of the impurity in its buckled state scales as $\sqrt{\frac{\gamma}{\gamma_c}-1}$.  Inset: Data on a log-log scale as a function of $(\gamma-\gamma_c)/\gamma_c$, clearly showing a slope of 1/2 close to the transition. Data shown are for a periodic hexagon with radius $R=15 a_0$, $\epsilon=0.1$, and $\gamma$ is changed by varying $\kappa$. 
}}
\end{figure}

We now focus on values of $\gamma \gtrsim \gamma_c$ just past the transition, such that $\gamma$ is not exceptionally large, the buckled profile is smooth, and the physics is more likely to be describable in terms of continuum elasticity \new{(we would not expect the prismatic profile's sharp lattice-scale features shown in Fig. \ref{collapse}(c) to be well suited to a continuum approach, for example)}. The height of the impurity in the transition region just above $\gamma_c$ is plotted in Fig. \ref{scaling}. As shown in the inset, the height grows as $\sqrt{(\gamma- \gamma_c)/\gamma_c}$. This behavior is reminiscent of the mean field behavior of the zero field ferromagnetic or antiferromagnetic Ising model, with $\gamma$ playing the role of temperature and the impurity height playing the role of an order parameter. We will introduce a theory that reproduces this result in Sec. \ref{ferro}.

The height profile of the buckled state close to the transition is also of interest, especially because it influences the interactions between dilations when multiple buckled impurities are present. When the vertical displacement caused by each impurity falls off quickly, as illustrated in Fig. \ref{collapse}(c) for $\gamma \to \infty$, well-separated impurities no longer influence each other. \new{This is also the case for dilations prior to buckling \cite{eshelby2}---for $\delta$-function impurities as in Eq. (\ref{metric}), the interaction energy for impurities at $\*r_1$ and $\*r_2$ in flat space goes as $\delta(\*r_1-\*r_2)$. However, at intermediate values of $\gamma \gtrsim \gamma_c$, as we will see, impurities do interact with each other, and the length scale with which vertical displacements decay then determines the interaction range of dilations on a lattice. }

\begin{figure}
\begin{center}
\includegraphics[width=\columnwidth]{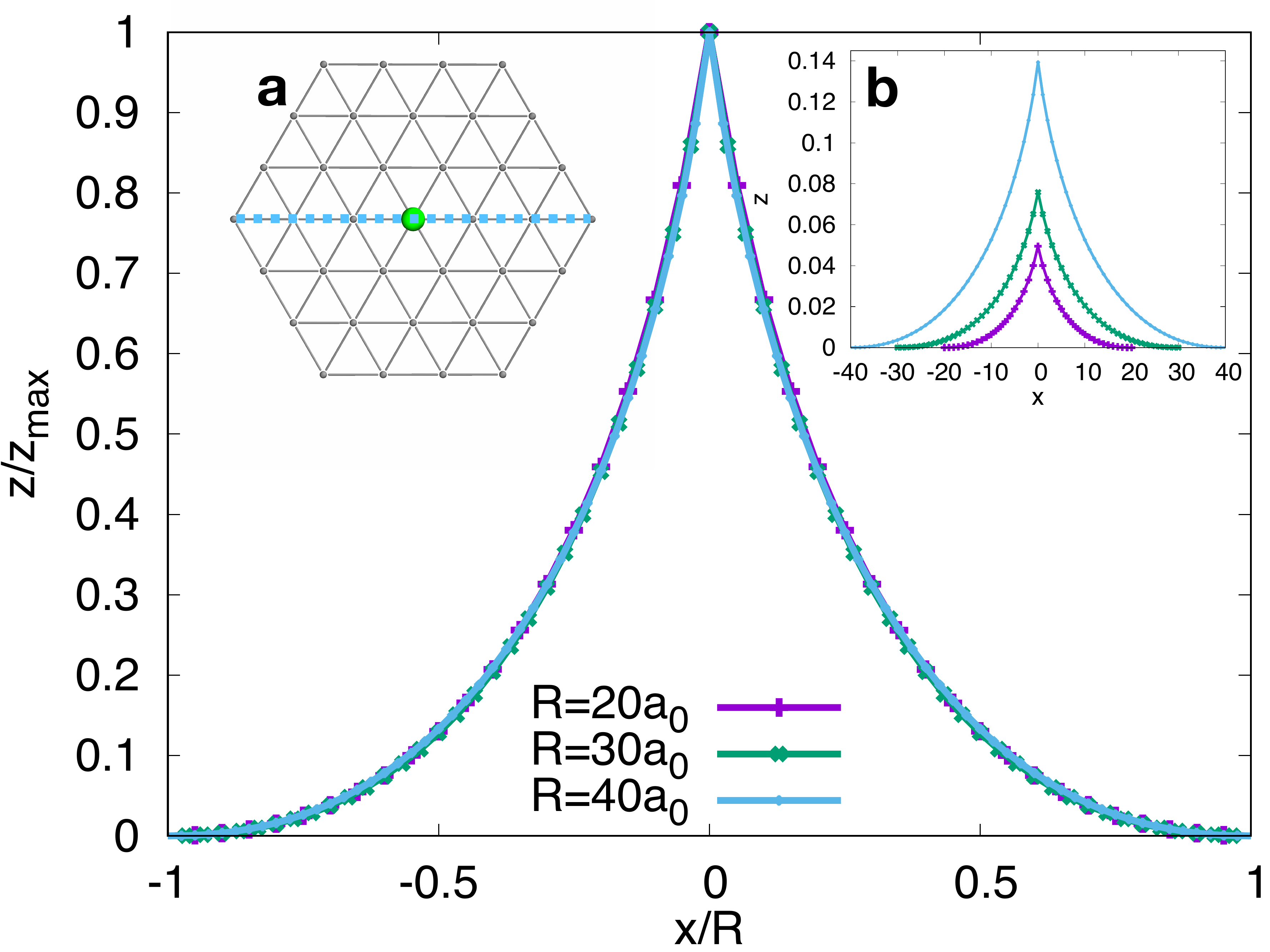}
\end{center}
\caption{{\label{height} Height profiles $z$ vs. $x$ for lattices of different sizes collapse when rescaled in the $z$ direction by the impurity height at the origin and rescaled in the $x$ direction by the system size, confirming the extensive nature of the deformation close to the buckling transition. Data are for $\gamma=\gamma_c+0.05$, with $\gamma_c$ approximated using simulations for each value of $R$ (see the discussion of Fig. \ref{vl} for details). (a) Height profiles are found by taking the height at each node along the dotted line, shown here for a lattice with $R=3a_0$. Results are similar if the height profiles are instead taken along a line at 30 degrees to the one pictured. (b) If we do not rescale $x$ and $z$, lattices of different membrane sizes have different height profiles.}}
\end{figure}

Remarkably, the dominant length scale close to the buckling transition with $\gamma \gtrsim \gamma_c$ appears to be the system size itself. At first sight, a dilation seems like a relatively minor perturbation to an elastic sheet (compared to a more extreme, topological lattice defect such as a disclination \cite{seung}), unlikely to have an effect on distant nodes. However, near the buckling transition, these systems prefer to distribute the dilation deformations globally and pay a penalty in stretching energy in order to avoid bending further. We show this effect in Fig. \ref{height}. The height of the impurity also increases with $R$ just past the buckling transition, as shown in Fig. \ref{height}(b); the height scales approximately as $\sqrt{R}$ when systems an equal amount past $\gamma_c$ are compared. We comment that the height of a right triangle with a hypotenuse of length $R+\epsilon a_0$ and base of length $R$ also has a height that scales as $\sqrt{R}$, and this may provide a crude approximation to the profile seen in Fig. \ref{height}. This system size dependence disappears as $\gamma \to \infty$. Thus, although the system size was not varied in Fig. \ref{collapse}, its variation would clearly have had an effect sufficiently close to the buckling transition. We note that the buckling threshold itself also has a correction due to finite system size that decays as $1/R$, which we will discuss in detail in Sec. \ref{ferro}. 

These observations are supported by recent work of \citet{oshri} who studied buckling in a closely related system in which a disk at the center of a larger circular region with free boundaries experiences dilational in-plane growth. This model can be thought of as an alternative continuum version of our discrete model for a single impurity with free boundaries, derived with a different coarse-graining procedure. \citet{oshri} also found that there is a near-threshold regime just past the buckling transition where height profiles are ``extensive" and the deformation spreads out over the whole system, and a far-from-threshold regime where the energy minimizing configurations are localized (analogous to our approach to the prismatic limit). Their study allowed for inclusions with a finite radius $R$, and they find that two dimensionless parameters control buckling, constructed with both the equivalent of $\gamma$ as well as $R$. For another paper related to this work, see \citet{efrati}.

To summarize the results of this section, increasing the dilation F{\"o}ppl-von K{\'a}rm{\'a}n number $\gamma$ with a single dilation embedded in a triangulated elastic sheet leads to buckling to a state with the potential for interactions with distant dilations near the transition. However, the strength of these interactions decays to zero as $\gamma \to \infty$ and the deformation becomes localized, as shown in the panels in Fig. \ref{collapse}.

\section{SINGLE IMPURITY WITH PERIODIC BOUNDARY CONDITIONS AS A FERROMAGNETIC ARRAY}
\label{ferro}
The preceding discussion relied heavily on observations emerging from a particular discretized ``tethered surface" model. We now supplement \new{these simulation results} for a single impurity with a nonlinear continuum theory that is able to predict the location of the buckling threshold and the scaling behavior of the impurity height near the transition. 

We first observe that our periodic boundary conditions are equivalent to considering an array of impurities. For the hexagonal system of radius $R$, the periodic image impurities are separated along the $(R/a_0,R/a_0)\equiv (n,n)$ direction [using the notation defined in Fig. \ref{lattice} and Fig. \ref{cappbc}(b)], and are all constrained to buckle identically on the same side of the host membrane. For comparison with later results, we call this configuration of buckled periodic images a ``ferromagnetic" array. Treating a single impurity system as a periodic array with a large defect spacing simplifies the analysis. 

\new{\subsection{Nonlinear continuum theory for a ferromagnetic array}}
We model our system with the elastic energy functional corresponding to the F{\"o}ppl-von K{\'a}rm{\'a}n equations with a contribution due to impurity defects from the generalization of Eq. (\ref{inplane}) \cite{landau, nelsondefects},

\begin{equation}
E= \int d^2 r \Big[\frac{\kappa}{2} \left(\nabla^2 f\right)^2+\mu u_{\alpha \beta}^2 +  \frac{1}{2}\lambda u_{\gamma \gamma}^2 - \left(\mu+ \lambda\right) \Omega_0 u_{\gamma \gamma}  c(\*r) \Big],
\label{energy}
\end{equation}
with 
\begin{align}
u_{\alpha \beta} &= \frac{1}{2} \left( \frac{\partial u_\alpha}{\partial x_\beta} + \frac{\partial u_\beta}{\partial x_\alpha}+ \frac{\partial f}{\partial x_{\alpha}} \frac{\partial f}{\partial x_\beta} \right),\\
&\equiv\frac{1}{2} \left( \frac{\partial u_\alpha}{\partial x_\beta} + \frac{\partial u_\beta}{\partial x_\alpha} \right) + A_{\alpha \beta}(\*r),
\label{strain}
\end{align}
and
\begin{equation}
c(\*r)= \sum_i \delta^2(\*r-\*r_i),
\label{cr}
\end{equation}
where $c(\*r)$ is the concentration of impurity defects--a sum of $\delta$ functions centered at regularly spaced impurity sites $\{\*r_i\}$. 

We now minimize the energy functional [Eq. (\ref{energy})] with respect to $u_\alpha$, the in-plane displacements, for a fixed function of the out-of-plane displacement field, $f(\*r)$. We will then substitute the minimizing displacements, $\bar{u}_\alpha$ back in to the original energy functional to arrive at an expression that depends only on $f(\*r)$ and the fixed impurity concentration $c(\*r)$. 

This procedure is related to that used at finite temperatures by \citet{radzihovsky} (among others \cite{wiese}) to integrate out in-plane phonons in a partition function. We will follow similar steps to isolate the effect of in-plane displacements and simplify the minimization.

We separate the strain tensor into $\*q=0$ and $\*q \ne 0$ Fourier modes and decompose $A_{\alpha \beta}(\*q)$ into longitudinal and transverse parts in terms of the functions $\phi_\alpha(\*q)$ and $\Phi(\*q)$. This gives 
\begin{align}
u_{\alpha \beta}(\*r)&= u_{\alpha \beta}^0 + A_{\alpha \beta}^0 \nonumber \\&+ \sum_{\*q \neq 0} \left( \frac{i}{2} \left[ q_\alpha u_\beta(\*q) + q_\beta u_\alpha(\*q) \right] +  A_{\alpha \beta} (\*q)\right) e^{i \*q \cdot \*r},\\
&= u_{\alpha \beta}^0+ A_{\alpha \beta}^0+ \sum_{\*q \neq 0} \bigg( \frac{i}{2} [ q_\alpha \left(u_\beta(\*q)+ \phi_\beta(\*q)\right) \nonumber \\ &+ q_\beta \big(u_\alpha(\*q)+ \phi_\alpha(\*q)\big) ] + P_{\alpha \beta}^T(\*q) \Phi (\*q)\bigg) e^{i \*q \cdot \*r},
\label{straintensor}
\end{align}
where $P_{\alpha \beta}^T(\*q)= \delta_{\alpha \beta} - \frac{q_\alpha q_{\beta}}{q^2}$ is the transverse projection operator. \new{Following \citet{nelsonbook}, $u_{\alpha \beta}^0$ corresponds to the uniform in-plane strains that are independent of $f$. The uniform strains that depend on $f$ are given by $A_{\alpha \beta}^0$.}

Upon inserting Eq. (\ref{straintensor}) into Eq. (\ref{energy}), and rewriting the resulting expression in terms of $w_\alpha(\*q)$, defined as
\begin{equation}
w_\alpha(\*q)=u_\alpha(\*q)+ \phi_\alpha(\*q) - \frac{\lambda i q_\alpha}{(2\mu + \lambda) q^2} \Phi (\*q) + \frac{ (\lambda + \mu)i q_\alpha}{(2 \mu + \lambda) q^2} \Omega_0 c (\*q),
\end{equation}
the energy becomes (to linear order in the extra impurity volume $\Omega_0$)
\begin{multline}
\frac{E}{A}= \mu \left(u_{\alpha \beta}^0+A_{\alpha \beta}^0\right)^2+\frac{\lambda}{2} \left(u_{\gamma \gamma}^0+A_{\gamma \gamma}^0\right)^2 \\- \left(\mu + \lambda\right) \Omega_0 \left(u_{\gamma \gamma}^0+A_{\gamma \gamma}^0\right) c(0) \\ + \frac{1}{2}\sum_{\*q \neq 0} \Big(\kappa q^4 |f(\*q)|^2 +  (\mu+\lambda)\left| \*q \cdot \*{w}(\*q) \right|^2+ \mu q^2\left|\*{w}(\*q) \right|^2\\+ Y\left| \Phi(\*q)\right|^2 -Y\Omega_0\Phi(\*q)c(-\*q)\Big),
\end{multline}
where $Y$ is the 2D Young's modulus $Y= \frac{4 \mu (\lambda + \mu)}{ 2 \mu + \lambda}$, and $A$ is the area of the unbuckled system, which also appears in our Fourier series convention $f(\*q)= \frac{1}{A} \int d^2r f(\*r)e^{-i \*q \cdot \*r}$. 

With this form of the total energy, we can now easily minimize over the quadratic dependence on in-plane displacements. Since the composite variable $w_\alpha(\*q)$ appears only as a magnitude squared, its contributions will be minimized when $w_\alpha(\*q)=0.$ Taking $w_\alpha(\*q)$ to be zero sets $\bar{u}_\alpha(\*q)$ for $\*q \neq 0$ in terms of $\*\phi, \Phi,$ and the Fourier transform of the impurity lattice. We can express $\*\phi$ and $\Phi$ in terms of the out-of-plane displacements,
\begin{equation}
\Phi(\*q)= P_{\alpha \beta}^T(\*q) P_{\alpha \beta}^T(\*q) \Phi(\*q) = P_{\alpha \beta}^T(\*q) A_{\alpha \beta}(\*q),
\end{equation} 
\begin{equation}
\phi_\alpha\*q)=\frac{-i}{q^2} \left[ 2 q_\beta A_{\alpha \beta}(\*q)- q_\alpha P_{\mu \nu}^L(\*q) A_{\mu \nu}(\*q) \right],
\end{equation} 
where
\begin{equation}
A_{\alpha \beta}(\*q)= \frac{1}{2A} \int d^2r \left(\frac{\partial f}{\partial x_\alpha}\frac{\partial f}{\partial x_\beta}\right) e^{-i \*q \cdot \*r},
\end{equation}
the Fourier transform of the nonlinear part of the strain tensor, defined above in Eq. (\ref{strain}), and $P_{\mu \nu}^L(\*q)=q_\mu q_\nu/q^2$ the longitudinal projection operator. \new{Note that none of these terms depend on in-plane displacements.  Since we are minimizing over $u_\alpha(\*q)$ assuming that $f$ is being held fixed, this means that we will be able to find a $\bar{u}_\alpha(\*q)$ that sets $w_\alpha(\*q)=0$, and we can fully determine our in-plane displacements. }

We can also use in-plane displacements to eliminate the energetic contribution from the $\*q=0$ mode by setting
\begin{equation}
\bar{u}_{\alpha \beta}^0 = -A_{\alpha \beta}^0+ \Omega_0 c(0) \delta_{\alpha \beta}.
\end{equation}
Since $c(\*q)= \frac{1}{A} \int d^2r c(\*r)e^{-i \*q \cdot \*r}$, $c(0)$ is simply the number density of impurity atoms, $M/A$. The $\*q=0$ mode is free to assume this value in our model because of our tension-free periodic boundary conditions. 

We can now write our energy minimized with respect to in-plane displacements in both Fourier and real space (neglecting terms of order $\Omega_0^2$) as
\begin{align}
E&=\frac{A}{2}\sum_{\*q \neq 0} \left(\kappa q^4 |f(\*q)|^2 + Y\left| \Phi(\*q)\right|^2 -Y\Omega_0\Phi(\*q)c(-\*q)\right),\\
&=\frac{1}{2}\int^\prime d^2r \bigg(\kappa \left(\nabla^2 f\right)^2 + Y\left(\frac{1}{2} P_{\alpha \beta}^T \partial_\alpha f \partial_\beta f\right) ^2 \nonumber \\&\hspace{9 em}-Y\frac{\Omega_0}{2}P_{\alpha \beta}^T \partial_\alpha f \partial_\beta f c(\*r)\bigg),
\label{energynophonons}
\end{align}
where the prime on the integral reminds us that the $\*q=0$ mode is excluded. \new{This energy has the same form as the effective free energy in \citet{radzihovsky} to linear order in $\Omega_0$. We echo their insight that the Laplacian of the term $\frac{1}{2} P_{\alpha \beta}^T \partial_\alpha f \partial_\beta f$ is approximately the Gaussian curvature, so the stretching energy is minimized when the Gaussian curvature of the dilation to be proportional to the Laplacian of a $\delta$ function. We note, however, subtleties associated with crushed vacancies \cite{jain}.}

We focus first on the terms quadratic in $f(\*q)$. Upon rewriting in terms of $f(\*q)$, the portion of the energy quadratic in $f(\*q)$ reads
\begin{align}
&\frac{E_2}{A}=\frac{1}{2} \sum_{\*q \neq 0} \kappa q^4 f(\*q)f(-\*q) \nonumber \\ &+\frac{Y \Omega_0}{4 v}\sum
\limits_{\substack{\*q^\prime+\*q^{\prime \prime}=\*q \neq 0 \\ \*G=-\*q \neq 0}}P_{\alpha \beta}^T(\*q^\prime + \*q^{\prime \prime}) q^\prime_\alpha q^{\prime \prime}_\beta f(\*q^\prime) f(\*q^{\prime \prime}) \delta_{\*G,-\*q^\prime-\*q^{\prime \prime}} ,
\label{quad}
\end{align}
\new{having specified $c(\*q)=\frac{1}{v} \sum_\*G \delta_{\*q,\*G}$, the Fourier transform of Eq. (\ref{cr}),} where $v$ is the real space area of the unit cell and $\*G$ is a reciprocal lattice vector, both corresponding to the \textit{impurity} superlattice. 

Since we require that the system is invariant under translations respecting the periodic boundary conditions, we expand $f(\*q)$, the out-of-plane displacement, in the corresponding set of superlattice reciprocal lattice vectors $\{\*G\}$, 
\begin{equation}
f(\*r)=\sum_{\*G} f(\*G)e^{i \*G \cdot \*r}
\label{gexpand}
\end{equation}
For impurities in a periodic $(n,n) =(R/a_0, R/a_0)$ array, the primitive vectors of the reciprocal impurity lattice are
\begin{align}
\label{recip1}
\*G_1&= -\frac{2 \pi}{3R} \*{\hat{x}} +\frac{2 \pi}{R\sqrt{3}} \*{\hat{y}} \equiv -\frac{g_0}{2}\*{\hat{x}} + \frac{g_0 \sqrt{3}}{2} \*{\hat{y}},\\
\*G_2&= \frac{2 \pi}{3R} \*{\hat{x}} +\frac{2 \pi}{R\sqrt{3}} \*{\hat{y}}\equiv \frac{g_0}{2}\*{\hat{x}} + \frac{g_0 \sqrt{3}}{2} \*{\hat{y}},
\label{recip2}
\end{align}
with 
\begin{equation}
g_0=\frac{4 \pi}{3R},
\end{equation}
the lattice spacing in reciprocal space, and a real space area of the unit cell
\begin{equation}
v=R^2 \frac{3 \sqrt{3}}{2}.
\end{equation}
We will also work with a third reciprocal lattice vector of the same magnitude,
\begin{equation}
\*G_3=\*G_2-\*G_1=\frac{4 \pi}{3R} \*{\hat{x}}= g_0 \*{\hat{x}}.
\end{equation}

\begin{figure}
\begin{center}
\includegraphics[width=\columnwidth]{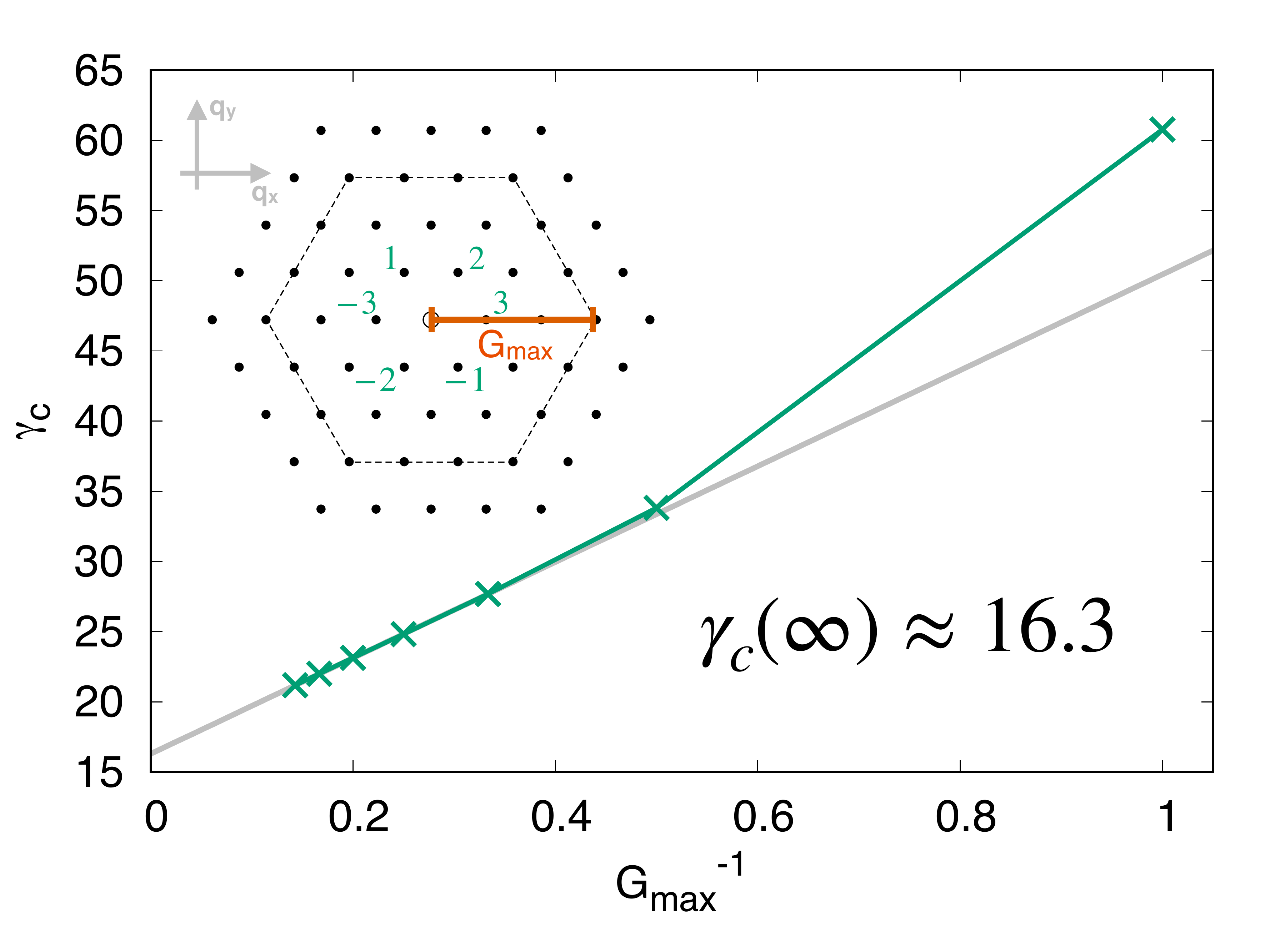}
\end{center}
\caption{{\label{bz} Variation of $\gamma_c$ with $G_{\text{max}}=n g_0$. When plotted as a function of $G_{\text{max}}^{-1}$, a linear extrapolation to the infinite system size limit gives $\gamma_c=16.3$. Results from various truncations of the Fourier expansion (\ref{gexpand}) used in our theory are shown for $n$ between 1 and 7, with $G_{\text{max}}$ measured in units of $g_0$. The green line connecting these points is a guide to the eye. Inset: Fourier modes used in the calculation when $G_{\text{max}}=3 g_0$ lie within or on the boundary of the dotted hexagon. Labeled wavevectors are the innermost ring of six included in the first calculation of $\gamma_c$ [Eq. (\ref{gamcfirst})].
}}
\end{figure}

We expect the lowest energy buckled states near the transition can be approximately described by the smallest reciprocal lattice vectors, corresponding to the longest wavelength deformations possible under our assumption of a periodic array. As a first approximation, we assume $f(\*r)$ is a linear combination of only the six smallest reciprocal lattice vectors,$\{\*G_1, \*G_{-1}, \*G_2, \*G_{-2}, \*G_3, \*G_{-3}\}$, with labels shown in the inset of Fig. \ref{bz}.

We now write the energy given this limited subspace for both $f(\*q)$ and $c(\*q)$. Since our displacements are real, we require $f^*(\*G_j)= f(-\*G_j)$. We impose this constraint and work in the six-dimensional subspace composed of the real and imaginary parts of $f(\*G_1)$, $f(\*G_2)$ and $f(\*G_3)$, and introduce the shorthand $\text{Re}[f(\*G_j)]=f_j^R$ and $\text{Im}[f(\*G_j)]=f_j^I$. 
We then express the energy in matrix form such that
\begin{equation}
\frac{E}{A}= H_{nm}f_n f_m,
\label{matrix}
\end{equation}
with
\[\mathbf{H}=
\begin{bmatrix}
    \alpha & -\Delta & -\Delta & 0  & 0 & 0 \\
    -\Delta & \alpha & -\Delta & 0  & 0 & 0 \\
    -\Delta & -\Delta & \alpha & 0  & 0& 0 \\
   0 & 0 & 0 & \alpha  & -\Delta & \Delta \\
  0 & 0 & 0 & -\Delta  & \alpha & -\Delta \\
   0 & 0 & 0 & \Delta  & -\Delta & \alpha \\
\end{bmatrix},
\hspace{2 em}
\mathbf{f}=
\begin{bmatrix}
    f^R_{1} \\
     f^R_{2} \\
     f^R_{3} \\
     f^I_{1} \\
     f^I_{2} \\
     f^I_{3} \\
\end{bmatrix},
\]
where $\alpha=\kappa g_0^4$ and $\Delta=\frac{3Y \Omega_0 g_0^2}{8 v}$. To determine the stability of the unbuckled state, we solve for the six eigenvalues and eigenvectors of $H_{nm}$. The eigenvalues are
\begin{align}
\bigg(\kappa g_0^4- \frac{3Y \Omega_0 g_0^2}{4 v}, & \kappa g_0^4- \frac{3Y \Omega_0 g_0^2}{8 v}, \nonumber \\ &\kappa g_0^4+ \frac{3Y \Omega_0 g_0^2}{8 v}, \kappa g_0^4+ \frac{3Y \Omega_0 g_0^2}{4 v} \bigg),
\label{evals}
\end{align}
where the second and third eigenvalue are doubly degenerate. Since all of the variables in Eq. (\ref{evals}) are positive, only the first two eigenvalues listed can attain negative values. The first eigenvalue will always give the lowest energy. We thus find that, \new{at this level of approximation,} our system becomes unstable to the corresponding buckling eigenvector provided $\gamma> \gamma_c$, where
\begin{equation}
\gamma_c\equiv\frac{Y \Omega_0}{\kappa}= \frac{4 v g_0^2}{3}= \frac{4}{3}\frac{8 \pi^2}{\sqrt{3}} \approx 61.
\label{gamcfirst}
\end{equation} 
The second eigenvalue becomes negative at $\gamma=2\gamma_c$. Note that the dilation F{\"o}ppl-von K{\'a}rm{\'a}n number $\gamma$ emerges naturally in this calculation, and all factors of $R$ cancel in the product $v g_0^2$. This calculation provides a rough estimate of the buckling threshold, but it is greater than that measured in simulations by approximately a factor of four. 

\begin{figure}
\begin{center}
\includegraphics[scale=0.22]{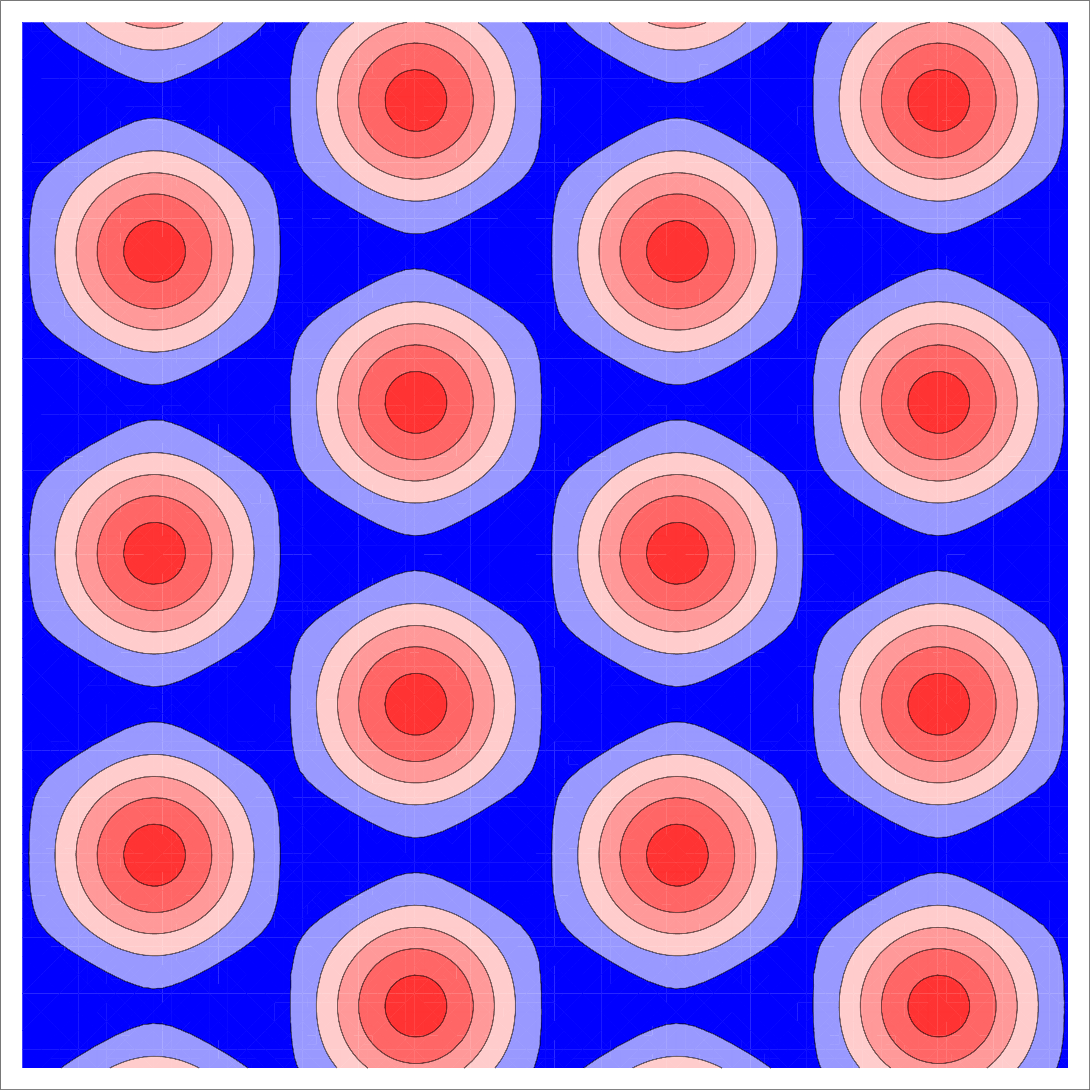}
\end{center}
\caption{{\label{m1} Contour plot of the first unstable eigenvector in our continuum elastic analysis of the buckling of a single dilation with periodic boundary conditions. The buckling amplitude is maximized at the centers of the red circles. Negative deflections $f(x,y)<0$ are blue, positive deflections are red. The magnitude is arbitrary, and the area displayed is $(6R)^2$. 
}}
\end{figure}

We now consider the eigenvector corresponding to the lowest energy eigenvalue (the first mode to go unstable). We define $m_1$ as its magnitude. This eigenvector gives the real space deformation
\begin{equation}
f(\*r)=\frac{2 m_1}{\sqrt{3}} \left[ \cos(\*G_1 \cdot \*r) + \cos\left(\*G_2 \cdot \*r \right) +\cos\left( \*G_3 \cdot \*r \right)\right],
\end{equation} 
pictured in Fig. \ref{m1}. Although at this level of approximation the eigenvector is independent of $\gamma$, this independence will not hold when more reciprocal lattice vectors are included in Eq. (\ref{gexpand}). 

We assume that just past the transition, the deformation $f(\*q)$ can be approximately described by this eigenvector. Contributions from the quartic stretching energy term can then easily be included. In general, the quartic term in the energy (\ref{energy}) is given by
\begin{multline}
E_4=\frac{YA}{8} \sum\limits_{\substack{ \*q_1+\*q_2 =\*q \neq 0 \\ \*q_3+\*q_4 =-\*q \neq 0}} P^T_{\alpha \beta}(\*q_1+\*q_2)q_{1\alpha} q_{2 \beta}
\\\times f(\*q_1)f(\*q_2)P^T_{\mu \nu}(\*q_3+\*q_4)q_{3 \mu} q_{4 \nu}f(\*q_3)f(\*q_4)\delta_{\sum_i \*q_i,0}
\end{multline}
Unlike the quadratic energy term, the quartic term is strictly positive, ensuring a finite value for the impurity height that minimizes the energy. The total energy as a function of $m_1$ is then 
\begin{equation}
\frac{E}{A}=\frac{\kappa g_0^4}{\gamma_c} \left( \gamma_c - \gamma \right) m_1^2 + \frac{3 Y g_0^4}{16} m_1^4,
\end{equation}
strikingly similar to the Landau free energy of the Ising model. We find the usual mean field ``critical exponent" $\beta=1/2$ when we minimize the energy with respect to $m_1$ when $\gamma > \gamma_c$.
\begin{equation}
m_1= \pm \sqrt{\frac{8 \kappa}{3 Y \gamma_c } \left( \gamma-\gamma_c\right)}=\pm\sqrt{\frac{8 \Omega_0}{3 \gamma } \left( \frac{\gamma}{\gamma_c}-1\right)}.
\end{equation}
This scaling behavior agrees with our data, shown in Fig. \ref{scaling}, and is consistent with prior work in real space by \citet{carraro}.

In fact, this result does not require assuming that only the buckling mode with magnitude $m_1$ is present: At this level of approximation we can prove that the two other eigenvectors that can have negative eigenvalues do not appear, even for $\gamma>2\gamma_c$. If we consider a height field composed of an arbitrary mixture of the first three eigenvectors, the energy to quartic order has a symmetry such that $m_2$ and $m_3$ always appear in the combination $m_2^2+m_3^2=\rho^2$, \new{and the energy can be mapped on to the Landau free energy of two competing Ising order parameters (see Ref. \cite{chaikin}, p. 181).} Given that $\frac{Y \Omega_0 g_0^2}{v}>0$, $\rho$ will always be zero, and the only phase transition in the system is the continuous transition in $m_1$ that we have already observed. 

\new{\subsection{Comparison between continuum theory and the discrete model}}
Although including only six reciprocal lattice vectors in the continuum calculation above allowed us to derive the scaling behavior we observe in our simulations, the calculated value of $\gamma_c$ [Eq. (\ref{gamcfirst})] is well above our numerical results, due, we believe, to our truncated basis in Fourier space. The energy minimizing structures we observe in simulations will surely have contributions from higher Fourier modes [see, for example, the short distance structure embodied in Fig. \ref{collapse}(a)]. We therefore repeat the calculation with higher Fourier modes and find that we arrive at a more accurate estimate.

Physically, as we raise the cutoff for the included Fourier modes, we allow for better resolution of the degrees of freedom embodied in the space between the impurity sites, while keeping the strength of the impurity and the distance between impurities the same. Systematically raising the maximum allowed $|\*G|=G_{max}$ in the expansion Eq. (\ref{gexpand}) is similar to measuring $\gamma_c$ in our discrete model for a single impurity with periodic boundaries as we progressively increase $R=n a_0$ for a fixed value of $a_0$. We have explored the agreement between theory and simulations by looking for trends in $\gamma_c$ as we increase both the number of included modes in the theory and the size of the system in the simulations. We do not expect quantitative agreement for small systems or few Fourier modes, since the discreteness of the host lattice has a large effect and the approximation of our impurities as $\delta$ function dilations in our theoretical treatment breaks down in this limit. Agreement between the theory and numerics should improve as the maximum reciprocal lattice vector and system size are increased and we approach the continuum limit.

It is helpful to examine this argument in more detail in terms of our discrete model. The lattice constant of the reciprocal lattice associated with the triangular mesh of the host lattice is $g_h=\frac{4 \pi}{\sqrt{3} a_0}$. However, the lattice constant of the impurity reciprocal lattice for an $(n,n)$ array is $g_0=\frac{4 \pi}{3 R} = \frac{4 \pi}{3 n a_0}$. Because the primitive vectors of these two lattices are at an angle of 30 degrees to one another, the first Brillouin zone of the host lattice is a hexagon of radius $G_{\text{max}}= \frac{4 \pi}{3 a_0}$ (see inset of Fig. \ref{bz}). When the radius of the real space lattice is increased from $R=n a_0$ to $R=(n+1)a_0$, the lattice spacing of the impurity reciprocal lattice shrinks. If we measure in units of the impurity reciprocal lattice spacing, the radius of the first Brilluoin zone also increases from $G_{\text{max}}=n g_0$ to $G_{\text{max}}=(n+1) g_0$. It therefore is reasonable to expect that the approach to the continuum limit will have the same scaling behavior for both the theory and simulations when plotted in the correct variables. 

\begin{figure}
\begin{center}
\includegraphics[width=\columnwidth]{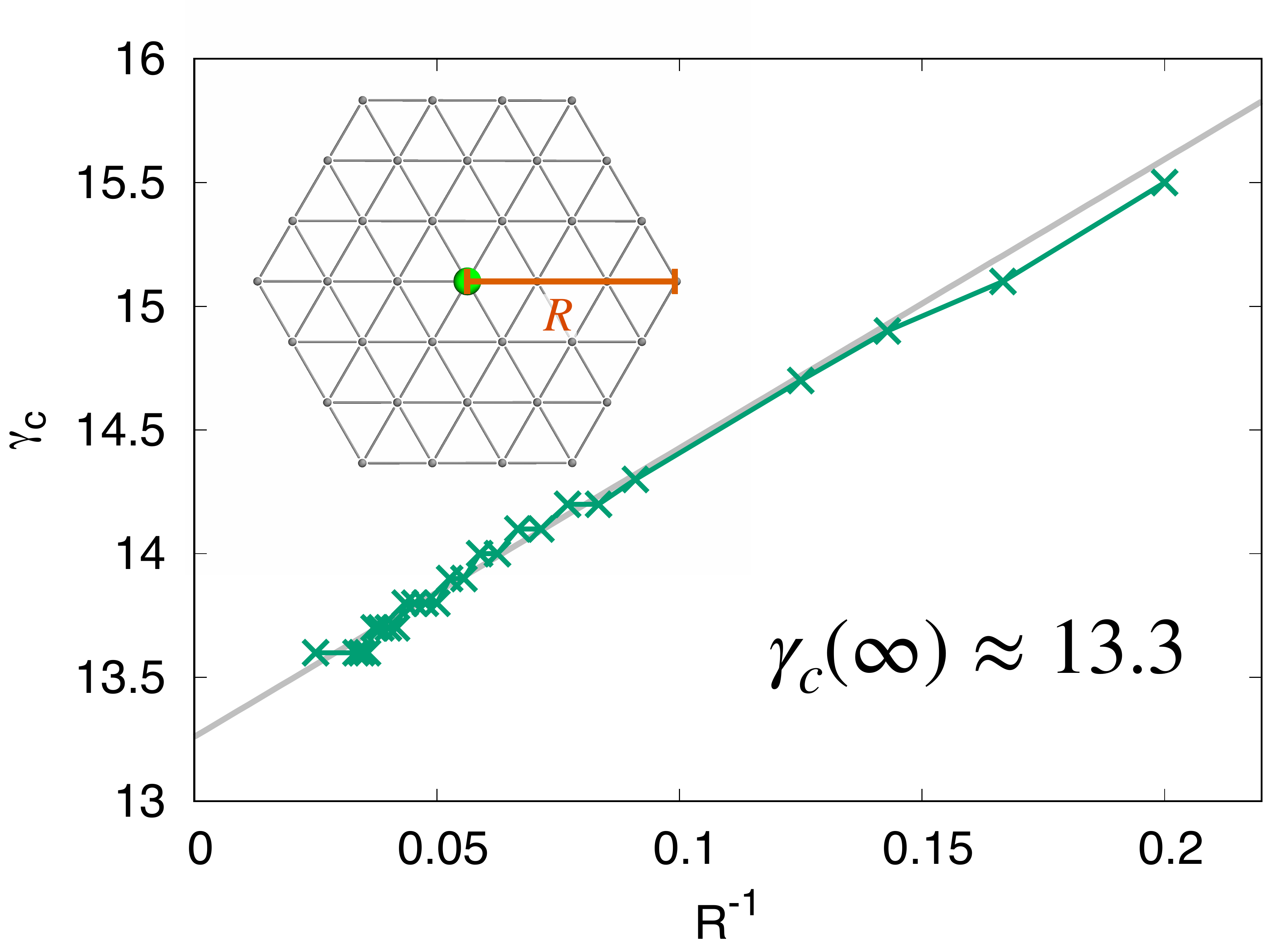}
\end{center}
\caption{{\label{vl} Variation of $\gamma_c$ from simulations with $R=n a_0$. When plotted as a function of $R^{-1}$ (units of $a_0^{-1}$), a linear extrapolation to the infinite system size limit gives $\gamma_c=13.3$. Data are shown for $n$ between $5$ and $40$, $\epsilon=0.05$, and $\gamma$ is changed by varying $\kappa$. 
Inset: Periodic unit cell of the defect superlattice for $R=3a_0$, \new{which would correspond to a $R^{-1}$ value to the right of those displayed.}
}}
\end{figure}

We test these ideas in Fig. \ref{bz} and Fig. \ref{vl}. In Fig. \ref{bz}, we calculate $\gamma_c$ by numerically finding the eigenvalues of the energy matrix as in Eq. (\ref{matrix}). In Fig. \ref{vl}, we estimate $\gamma_c$ for the discrete model by varying $\gamma$ with a resolution of $0.1$. As before, for a given value of $\gamma$, we displace the impurity node at the origin a fixed amount in the positive $z$ direction and minimize the energy. We compare this energy to the energy of the system with the same dilation when it is minimized in flat space. We set $\gamma_c$ to be the highest value of $\gamma$ for which the energy of the system that relaxes in three dimensions is equal (at our level of numerical precision) to the energy of the system that relaxes in two dimensions. For all values of $\gamma$ greater than this $\gamma_c$, the energy of the system that is allowed to buckle is lower.

The values of $\gamma_c$ in Fig. \ref{bz} and Fig. \ref{vl} approach a limit as we increase $R$ and $G_{\text{max}}$. We can estimate the infinite system size value of $\gamma_c$ by extrapolating $1/R$ and $1/G_{\text{max}}$ to $0$. This extrapolation allows us to predict that $\gamma_c(\infty) \approx 16.3$ from the Fourier space theory (Fig. \ref{bz}) and $\gamma_c(\infty)\approx 13.3$ from the simulations (Fig. \ref{vl}). We seem to have approximate agreement in the limit in which continuum theory should apply, \new{with a small discrepancy that is at least in part due to terms of order $\Omega_0^2$ that were neglected (we find that shrinking $\epsilon$ in simulations leads to small increases in $\gamma_c$).} Furthermore, as hoped, the scaling behavior is the same: $\gamma_c$ appears to be a linear function when plotted as a function of $1/R$ and $1/G_{\text{max}} $. A linear dependence on these quantities is plausible, since we expect corrections to the continuum limit to scale as the ratio of the hexagon perimeter to the hexagon area in both real and reciprocal space.

\section{SEARCHING FOR THE GROUND STATE OF AN IMPURITY ARRAY}
\label{groundstatesec}
\begin{figure}
\begin{center}
\includegraphics[scale=0.28]{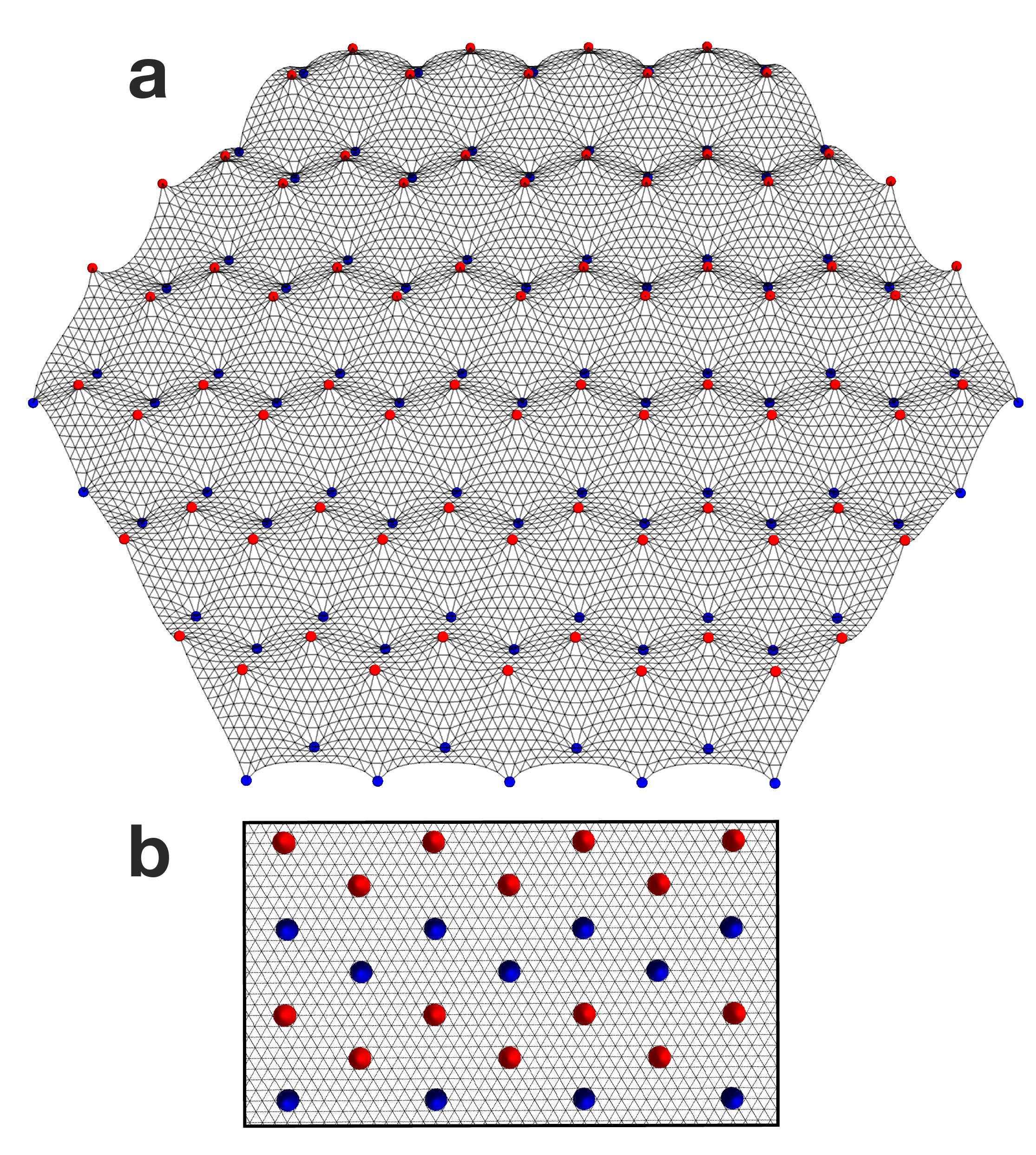}
\end{center}
\caption{{\label{zigzag} The buckled zigzag state for a $(4,4)$ array of dilations in a hexagonal periodic domain of size $R=48a_0$ at $\gamma=17.9$, our conjectured ground state for the triangular host lattice. (a) Zigzag state viewed in perspective, with the vertical displacements magnified by a factor of 10 for clarity. (b) Top-down view of the zigzag state with impurities that have buckled up shown in red, and impurities that have buckled down shown in blue. 
}}
\end{figure}

\new{In this section, we make an analogy between buckled impurities and Ising spins that provides seven candidate ground-state configurations for a buckled impurity array, and test these candidates by observing their buckling transition and measuring the energy. We find that a zigzag state (Fig. \ref{zigzag}) has the lowest energy by a small amount, which is consistent with the nonlinear theory introduced in Sec. \ref{ferro}.}

Past the buckling transition, each dilation can buckle either up or down out of the plane, possibly influenced by interactions with neighboring impurities. This bistability gives us a complex energy landscape with many metastable states. Although phenomena such as phase transitions at finite temperatures are quite interesting (see Sec. \ref{discussion}), we focus here on the ground state (or ground states) of the system at $T=0$. Understanding the ground state will provide insight into how the system organizes, and is a starting point for future investigations of fluctuations among the many metastable states at nonzero temperatures, where entropy can play an important role. 

However, determining which configuration of an interacting triangular superlattice of possibly buckled dilations has the lowest energy is a challenging problem. If we consider states characterized by their up and down patterns of $M$ buckled impurities at fixed $\gamma>\gamma_c$, there are of order $2^M$ candidate ground states to test. Since we do not have the computational resources to test all of these states for large $M$, it is difficult to prove unambiguously which state has the lowest energy. Instead, we will use physical reasoning, simulations, and calculations to conjecture a likely ground state for $\gamma>\gamma_c$. We suspect, but cannot prove, that the buckling pattern of up and down dilations with the lowest energy for $\gamma \gtrsim \gamma_c$ will remain the pattern with the lowest energy for $\gamma \gg \gamma_c$. 

We will continue to probe the system by varying $\gamma$ and the number of mesh spacings of the host lattice separating impurities. Although we will focus on $(n,n)$ arrays, two other array families are also interesting. $(0,n)$ arrays buckle in ways that depend more strongly on $n$ and other microscopic details of the underlying host lattice. \textit{Chiral} arrays, which have $(n,m)$ with $n \neq 0$, $m \neq 0$, and $n \neq m$ interact more strongly with our planar boundary conditions. The $(n,n)$ arrays we study in detail have a smoother approach to the continuum limit, and allow us to directly apply lessons from the single impurity case. 

\new{\subsection{Analogy with the Ising model}}
Even after restricting ourselves to $(n,n)$ arrays, the parameter space of the problem is massive. The problem becomes more tractable if we assume that the ground state is determined by pairwise interactions between buckled impurities. This approximation allows us to be guided by previous work on the Ising model on a fixed triangular lattice in flat space. \new{Drawing a connection to the Ising model has aided in the understanding of mechanical systems with discrete degrees of freedom before; see Refs. \cite{spinice} and \cite{han} for particularly relevant examples.}

\new{We can re-express our impurity array as a spinlike model by associating the impurity out-of-plane displacements with Ising spins.} If an impurity buckles up, we assign it to be spin up, and \textit{vice versa}. Sufficiently close to the buckling transition, the distortions caused by two nearby buckled impurities will overlap. The interaction energy will differ depending on whether the impurities are buckled in the same direction or different directions (i.e., if the ``spins" are aligned or antialigned). Depending on whether the energy is lower for the aligned or anti-aligned configuration, the local interaction is ferromagnetic or antiferromagnetic, respectively \cite{carraro}. 

Since the interactions between impurities can be long range \new{when the system is close to the buckling transition,} we allow the impurity ``spins" to have interactions with not only their nearest-neighbor ``spins," but also with next-nearest and third-nearest neighbors. \new{Including longer-range interactions is physically motivated for $\gamma \gtrsim \gamma_c$, and will also lift the ground-state degeneracy due to geometric frustration if the couplings are antiferromagnetic \cite{tanaka}.}

There are seven ground states possible for an Ising model on a rigid triangular lattice with up to third-nearest-neighbor interactions \cite{tanaka}. \new{Which of the seven states is the ground state depends on the sign and magnitude of the three spin coupling constants. We translate the seven spin configurations into buckled impurity array states; see Appendix C for top-down views of all seven.} 

\new{We assume that the ground state of our buckled impurity system corresponds to one of these seven candidate configurations, and test all of them.} Since the energy minimization algorithm used with our discrete model finds the nearest local minimum, we can probe metastable states by initializing simulations with small positive and negative vertical displacements on the impurity atoms in the desired Ising spin pattern \new{(as we did for the single impurity in Sec. \ref{single}). We confirm that impurity ``spins" do not flip during the energy minimization process, and compare the energies and impurity heights of the minimized configurations. }

\begin{figure}
\begin{center}
\includegraphics[width=\columnwidth]{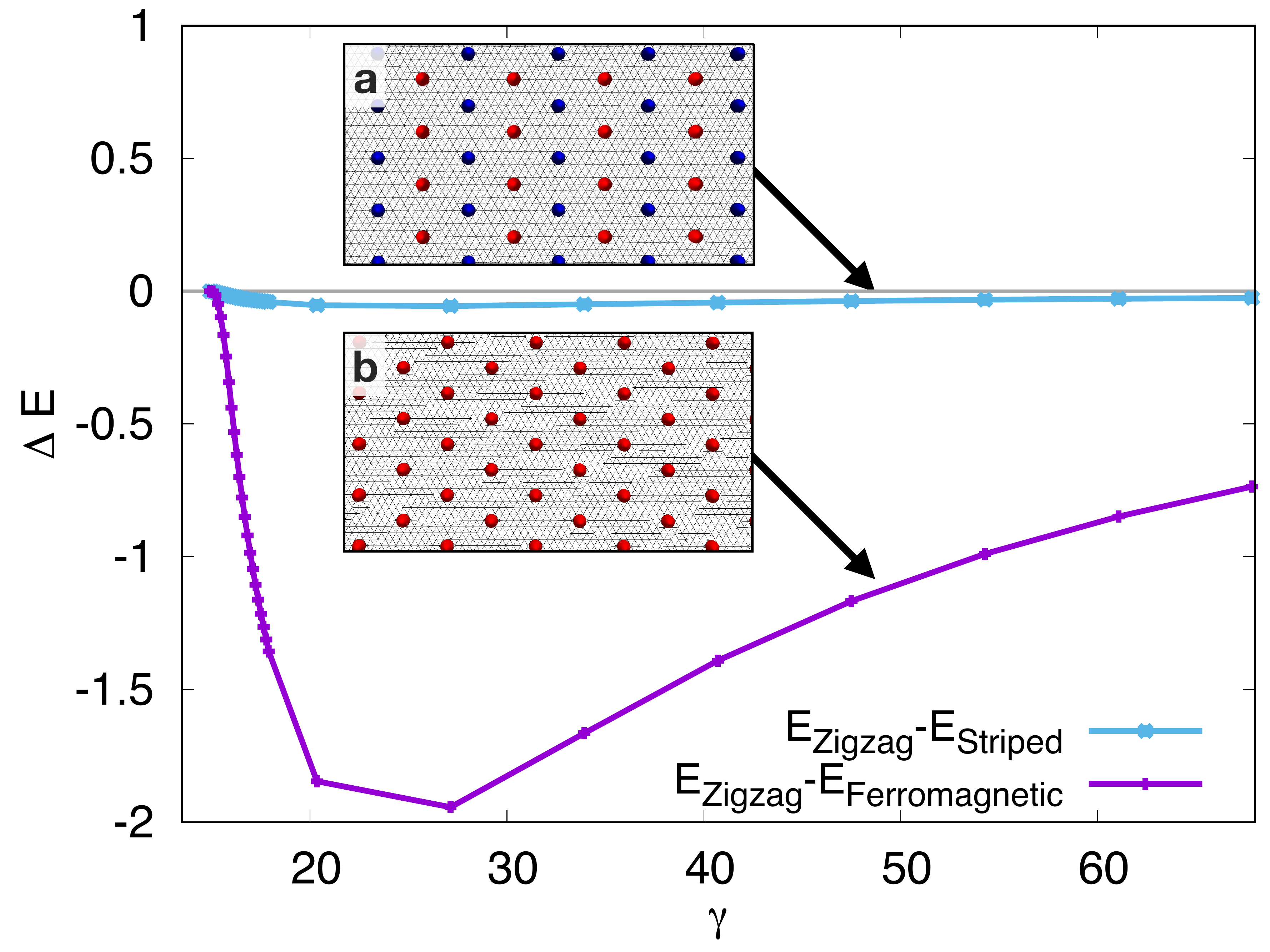}
\end{center}
\caption{{\label{dE} Difference in energy between two metastable states, the striped state (a) and the ferromagnetic state (b), and our conjectured zigzag ground state (pictured in Fig. \ref{zigzag}) for $\gamma$ larger than $\gamma_c$ of the zigzag state. All data are for $(4,4)$ arrays embedded in a hexagon with $R=96a_0$, $\epsilon=0.1$. The dilation F{\"o}ppl-von K{\'a}rm{\'a}n number $\gamma$ is changed by varying $\kappa$. 
}}
\end{figure}

We find, out of the seven Ising candidate states, the ``zigzag state" shown in Fig. \ref{zigzag} is the first state to buckle as $\gamma$ is increased, and has the lowest energy for all values of $\gamma$ tested once buckling has occurred. We plot the difference between the energy of the zigzag state and two other states in Fig. \ref{dE}. In Fig. \ref{dE}, and the rest of this work, we show results only for three states of interest: the conjectured zigzag ground state, the striped state, which is close in energy to the ground state and pictured in Fig. \ref{dE}(a), and the ``ferromagnetic" state, which has the highest energy of all metastable states we measured and is pictured in Fig. \ref{dE}(b).

The finding that the zigzag state has the lowest energy, closely followed by the striped state, suggests that all pairwise interactions are antiferromagnetic, with the strength of the interaction falling off with distance. \new{Note that in an antiferromagnetic spin system that has only nearest-neighbor interactions, the zigzag and striped spin states have the same energy-- each spin has four favorable and two unfavorable interactions with neighbors. Longer range interactions (such as those that appear naturally in our spinlike impurity arrays) break the degeneracy of the ground state.}

The conclusion that pairwise interactions are antiferromagnetic is supported by measurements of the energy of isolated pairs of up-up and up-down buckled impurities as the separation is varied. The strength of the antiferromagnetic interactions goes to zero as the separation between impurities is taken to infinity. These interactions also vanish in the prismatic limit $\gamma \to \infty$. 

\new{In a real space continuum treatment of dilations, we also find that the bending energy favors long-range antiferromagnetic interactions \cite{carraro}. As in Sec. \ref{single}, we estimate the bending energy of two nearby aligned or antialigned impurities by assuming the height profiles are Gaussian. However, we now allow $\sigma$, which measures the size of the buckled region, to be greater than the impurity core size since we will not require the stretching energy to be negligible. }

The superimposed Gaussian height profile,
\begin{equation}
f(x,y)=  H_0 \left(e^{-\left((x - d)^2 + y^2 \right)/2\sigma^2}\pm e^{-\left((x + d)^2 + y^2 \right)/2\sigma^2} \right),
\label{2gauss}
\end{equation}
gives us a bending energy of the form
\begin{equation}
E_b= \frac{\pi \kappa H_0^2} {\sigma^6} \left( 2 \sigma^4 \pm e^\frac{-d^2}{\sigma^2} (d^4- 4 d^2 \sigma^2 + 2 \sigma^4) \right).
\end{equation}
The bending energy difference between the antialigned (antiferromagnetic) and aligned (ferromagnetic) puckers is 
\begin{equation}
E_{+-}-E_{++}= -\frac{2 \pi \kappa H_0^2}{\sigma^2} e^{-\frac{d^2}{\sigma^2}}\left(\left(\left(\frac{d}{\sigma}\right)^2- 2\right)^2 - 2 \right),
\label{gaussbend}
\end{equation}
and is negative when $d/\sigma> \sqrt{2+\sqrt{2}} \approx 1.85.$ \new{Provided we can neglect the difference in stretching energy for the antiferromagnetic and ferromagnetic puckers, there are antiferromagnetic pairwise interactions for impurities with separations somewhat larger than the buckling size.} These conclusions once again hold if a more realistic height profile with an exponential decay is used (see Appendix A).

\new{This argument also seems to suggests the possibility of short-range ferromagnetic interactions, when impurity centers are very close. A more detailed analysis would be necessary to make this claim, since the precise form of the height profiles is more relevant in this case.} However, we did observe ferromagnetic pairwise interactions in a $(2,0)$ array, consistent with the behavior observed by \citet{carraro}, although the impurities are so close together in this case that the continuum description cannot be straightforwardly applied. We have avoided this regime in our simulations. 

\begin{figure}
\begin{center}
\includegraphics[width=\columnwidth]{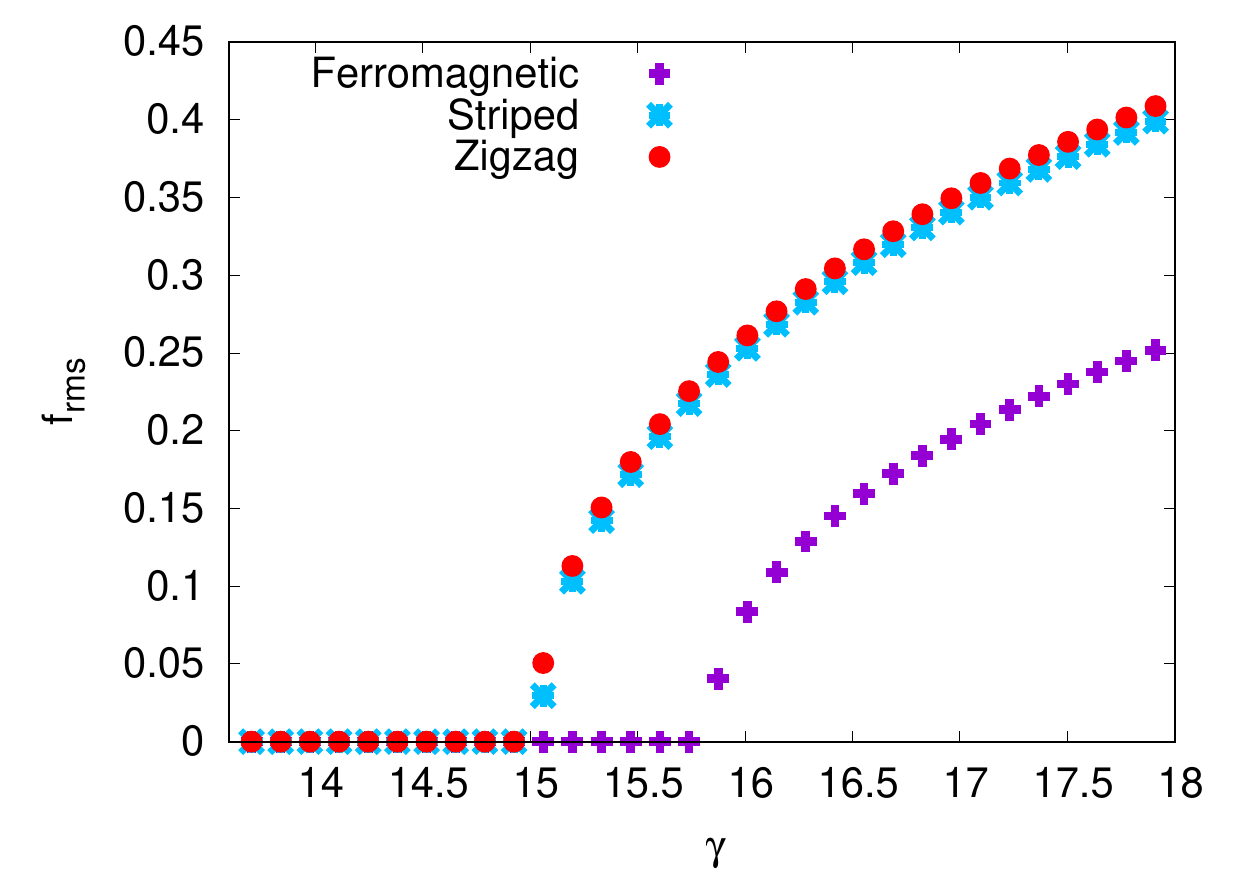}
\end{center}
\caption{{\label{buckle} Root-mean-square height at impurity sites (as a measure of the magnitude of up and down buckling) versus $\gamma$. The zigzag state buckles first, followed closely by the striped state, and then the ferromagnetic state. Data are for $(4,4)$ arrays with $R=96a_0$, $\epsilon=0.1$, and $\gamma$ is changed by varying $\kappa$.  
}}
\end{figure}

In addition to measuring the energy of various states containing many dilations, we can measure the buckling threshold. As seen in Fig. \ref{buckle}, the zigzag state buckles first. This observation is consistent with a zigzag ground state, since the buckling transition is defined as the point at which a new configuration attains an energy lower than the flat-state energy. It also means we can use our calculation of $\gamma_c$ to better understand ground-state behavior. As we now show, the analysis presented to study a single impurity as a ferromagnetic array readily generalizes to other types of deformations. 

\new{\subsection{Comparison between continuum theory and the discrete model}}
To proceed, we determine the subspace of wavevectors compatible with the periodicity of the zigzag and striped states. We can predict nonzero $f(\*q)$ modes by, in the language of solid state physics \cite{am}, looking for extinctions in the structure factor of the impurity superlattice. To simplify our analysis, we initially assume that each impurity has a height of either $1$ or $-1$, and all nonimpurity sites have a height of zero. Under this assumption, we can find the Fourier transform of our height field by treating up and down impurities as two different types of atoms with atomic form factors of $1$ and $-1$, respectively.

\begin{figure}
\begin{center}
\includegraphics[width=\columnwidth]{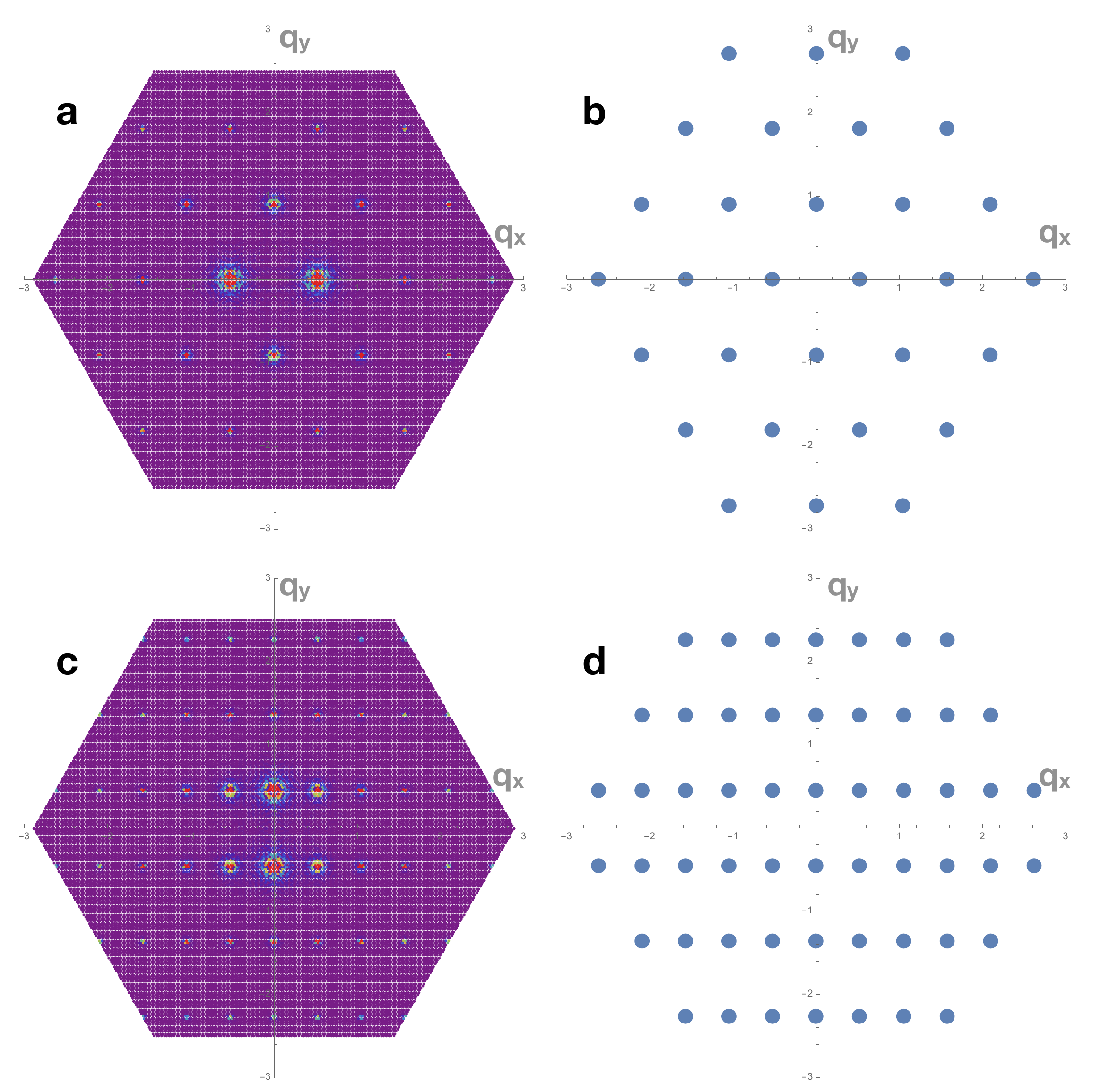}
\end{center}
\caption{{\label{ft} (a) Fourier modes for the height field computed with Eq. (\ref{verify}) for a striped configuration on a $(4,4)$ impurity array with $R=96a_0$, $\gamma\approx 68$. (b) Allowed wavevectors for a striped $(4,4)$ impurity array using Eq. (\ref{striped}). (c) Fourier modes for the zigzag state under the same conditions as in (a). (d) Allowed wavevectors for the zigzag state as in (b) using Eq. (\ref{zigzagstruc}).
}}
\end{figure}

We first describe a unit cell for the impurity height field with a basis.  For the striped state, the lattice vectors
\begin{equation}
\*a_1=3na_0 \hat{\*x}, \hspace{3 em} \*a_2 = \sqrt{3} n a_0 \hat{\*y},
\end{equation}
with the basis
\begin{equation}
\*d_1=0, \hspace{3 em} \*d_2=\frac{3 n a_0}{2} \hat{\*x} + \frac{\sqrt{3} n a_0}{2} \hat{\*y},
\end{equation}
give the appropriate description, with $\*d_1$ and $\*d_2$ corresponding to impurities that buckle up and down respectively. 
The reciprocal lattice vectors describing this deformation are
\begin{equation}
\*G(b_1,b_2)= b_1 \frac{2 \pi}{3 n a_0}\hat{\*x} + b_2 \frac{2 \pi}{\sqrt{3} n a_0}\hat{\*y},
\end{equation}
where $b_1$ and $b_2$ are integers, whose amplitude is modulated by the geometrical structure factor
\begin{equation}
S_{\*G}= \sum_{i}f_{\*d_i}(\*G)e^{i \*G \cdot \*d_i}=f_{\*d_1}(\*G)+f_{\*d_2}(\*G)e^{i\*G \cdot \*d_2}.
\end{equation}
If we rescale the atomic form factors $f_{\*d_i}(\*G)$ to unity, corresponding to identical atoms, we regain the reciprocal lattice described by Eqs. (\ref{recip1}) and (\ref{recip2}). However, the choice $f_{\*d_1}(\*G)=1$ and $f_{\*d_2}(\*G)=-1$, leads to
\begin{equation}
S_\*G = 1-\exp\left(i \pi (b_1+b_2)\right).
\label{striped}
\end{equation}
If $b_1+b_2$ is odd, the reciprocal vector characterized by $b_1$ and $b_2$ will survive. If $b_1+b_2$ is even, $S_\*G$ will be zero, and this wave vector will not show up in the Fourier transform of the striped state. This rule leads to the lattice shown in Fig. \ref{ft}(b).

For the zigzag state, the lattice vectors
\begin{equation}
\*a_1=3na_0 \hat{\*x}, \hspace{3 em} \*a_2 = 2\sqrt{3} n a_0 \hat{\*y},
\end{equation}
with the basis
\begin{align}
&\*d_1=0, \hspace{4 em} \*d_2=\frac{3 n a_0}{2}  \hat{\*x} + \frac{\sqrt{3} n a_0}{2} \hat{\*y}, \hspace{1 em}\nonumber\\ &\*d_3=\sqrt{3} n a_0 \hat{\*y}, \hspace{1 em} \*d_4=\frac{3 n a_0}{2}  \hat{\*x} + \frac{3\sqrt{3} n a_0}{2} \hat{\*y},
\end{align}
give the appropriate description, with dilations at $\*d_1$ and $\*d_4$ buckling up and those at $\*d_2$ and $\*d_3$ buckling down. 
The corresponding reciprocal lattice vectors
\begin{equation}
\*G(b_1,b_2)= b_1 \frac{2 \pi}{3 n a_0}\hat{\*x} + b_2 \frac{\pi}{\sqrt{3} n a_0}\hat{\*y},
\end{equation}
lead to a structure factor of
\begin{equation}
S_{\*G}= 1-e^{i\pi\left( b_1+\frac{b_2}{2}\right)}-e^{i\pi b_2}+e^{i\pi\left( b_1+\frac{3 b_2}{2}\right)},
\label{zigzagstruc}
\end{equation}
which gives the lattice shown in Fig. \ref{ft}(d).

To verify that we have identified the appropriate components of $f(\*q)$, we plot the intensity that would result from a scattering experiment,
\begin{equation}
I(\*q)=\frac{1}{N}\left| \sum_{\*r_i} f(\*r_i) e^{-i \*q \cdot \*r_i} \right|^2,
\label{verify}
\end{equation}
using simulation data of the states, where $\*r_i$ is the $(x,y)$ location of node $i$ on the lattice (both impurity and nonimpurity sites included, $N$ sites in total), and $f(\*r_i)$ is the height of that node. These results are shown in Fig. \ref{ft}(a) and Fig. \ref{ft}(c) for the striped and zigzag state, respectively. 
 
\begin{figure}
\begin{center}
\includegraphics[width=\columnwidth]{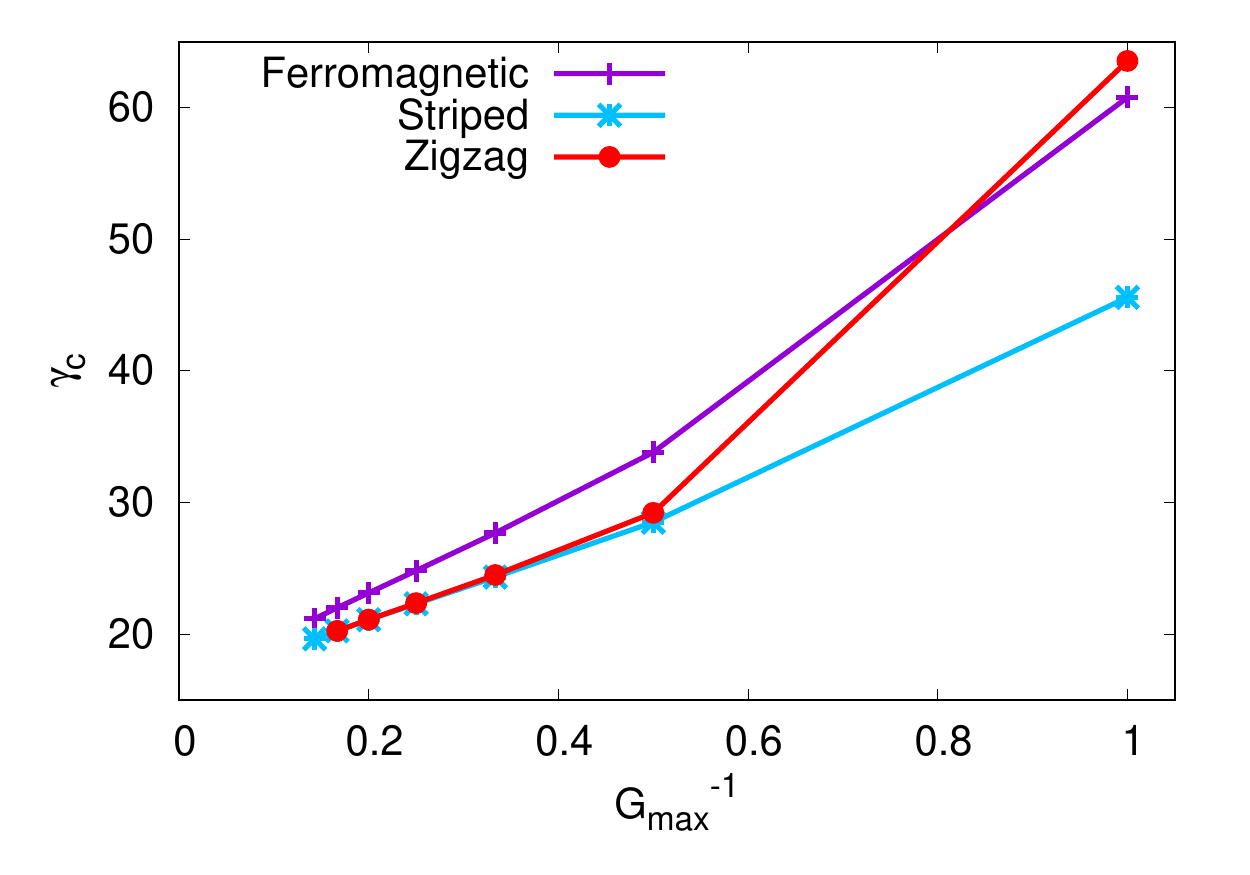}
\end{center}
\caption{{\label{compare} Variation of $\gamma_c$ with $G_{\text{max}}=n g_0$ for the ferromagnetic state (same values as in Fig. \ref{bz}), striped state, and zigzag state. Results from various truncations of the Fourier expansion Eq. (\ref{gexpand}) used in our theory are shown for $n$ between 1 and 7, with $G_{\text{max}}$ measured in units of $g_0$.
}}
\end{figure}

Calculating $\gamma_c$ as before for each of these structures in Fig. \ref{compare}, we find that the analysis correctly predicts that the striped and zigzag states buckle before the ferromagnetic state when a reasonable number of Fourier modes are included. We truncate Fourier space for each calculation such that the magnitude of the largest superlattice reciprocal lattice vector included in the calculation [see Eq. (\ref{quad})] is greater than or equal to the magnitude of the largest Fourier space component of $f$. Upon extrapolating to $G_{\text{max}}^{-1}=0$ (corresponding to an infinitely fine mesh for the host lattice of our dilations) and using a linear fit that neglects the two points with the fewest Fourier modes included, we find an estimate of $\gamma_c(\infty)=16.3$ for the ferromagnetic state, $\gamma_c(\infty)=16.2$ for the striped state, and $\gamma_c(\infty)=16.0$ for the zigzag state.

\begin{figure}
\begin{center}
\includegraphics[width=\columnwidth]{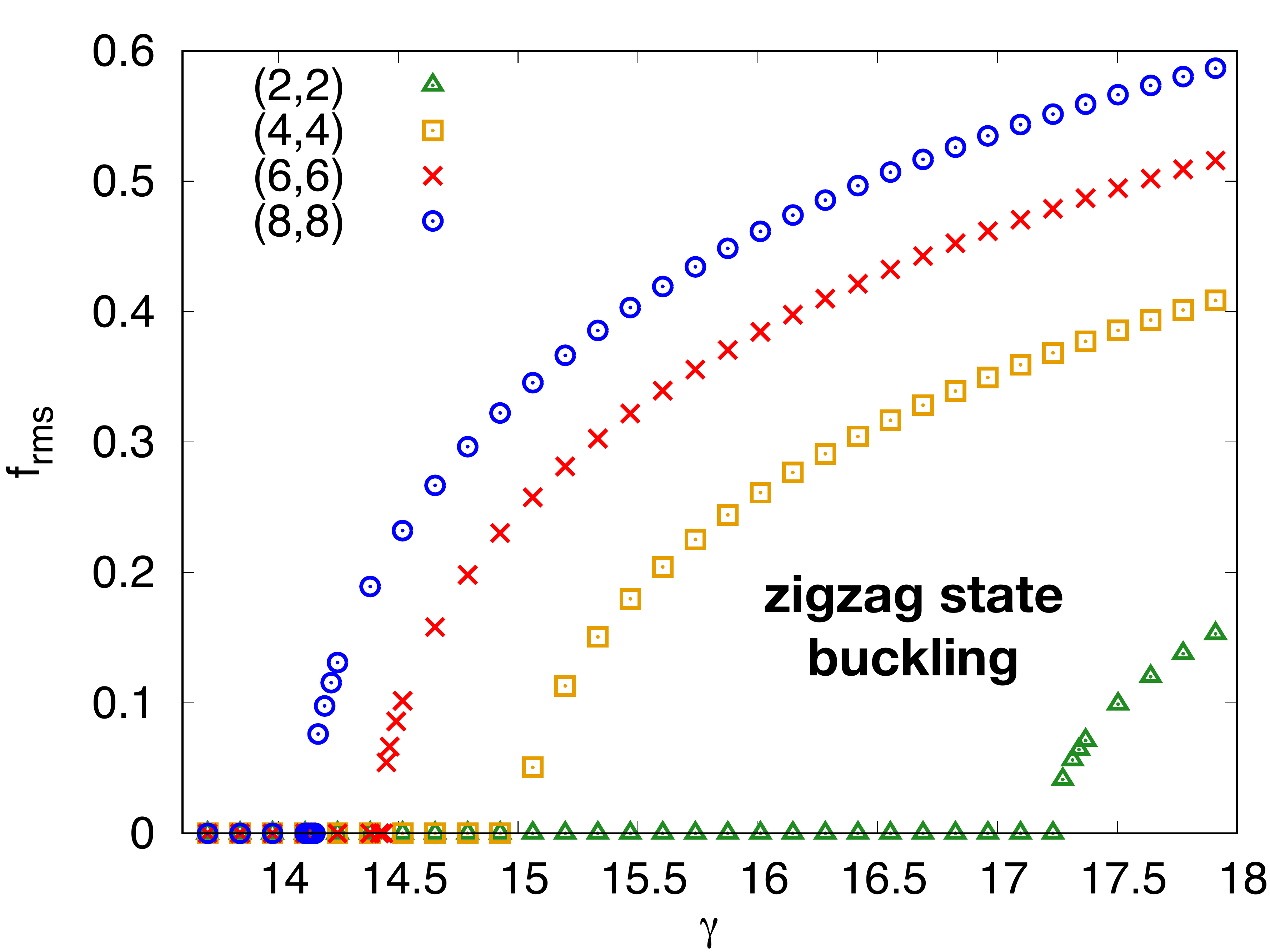}
\end{center}
\caption{{\label{zigzagR} Root-mean-square impurity height (as in Fig. \ref{buckle}) for the zigzag state in our simulations as a function of $\gamma$ with increasing impurity separation $(n,n)$. Data are for separations $n=2,4,6,$ and $8$ in a hexagonal domain of size $R=96a_0$ with $\epsilon=0.1$, and $\gamma$ is changed by varying $\kappa$. 
}}
\end{figure}

We can also see the \new{effect of increasing the number of mesh points between impurities} in our simulations, illustrated in Fig. \ref{zigzagR} for the zigzag state. As we increase the separation between impurities embedded in $(n,n)$ lattices, effectively increasing the number of wavevectors that can be included, the buckling threshold decreases in a manner consistent with a smooth approach to a limiting $\gamma_c(n)$ as $n \to \infty$. 

\section{EXTENSION TO A SQUARE LATTICE}
\label{square}
\new{In this section, we introduce and characterize a discrete model with a square host lattice. We find similar results for single impurity buckling as in Sec. \ref{single}, and conjecture a checkerboard ground state for impurity arrays (see Fig. \ref{checker} below) using the same methods as Sec. \ref{groundstatesec}. The energy gap between the checkerboard state and other square lattice states tested, as well as the difference in the buckling threshold, is much greater than what was observed with the triangular host lattice model.}

Explorations of the ground state for a triangular lattice of impurities are complicated by geometric frustration, a familiar difficulty in Ising-like systems \cite{tanaka}. Because pairwise interactions between neighboring dilations are antiferromagnetic, longer-range interactions are necessary to specify the ground state. A natural question to ask is: How do our results change for impurity buckling in a geometry where the ground state of, say, an antiferromagnetic configuration is not frustrated?

\begin{figure}
\begin{center}
\includegraphics[width=\columnwidth]{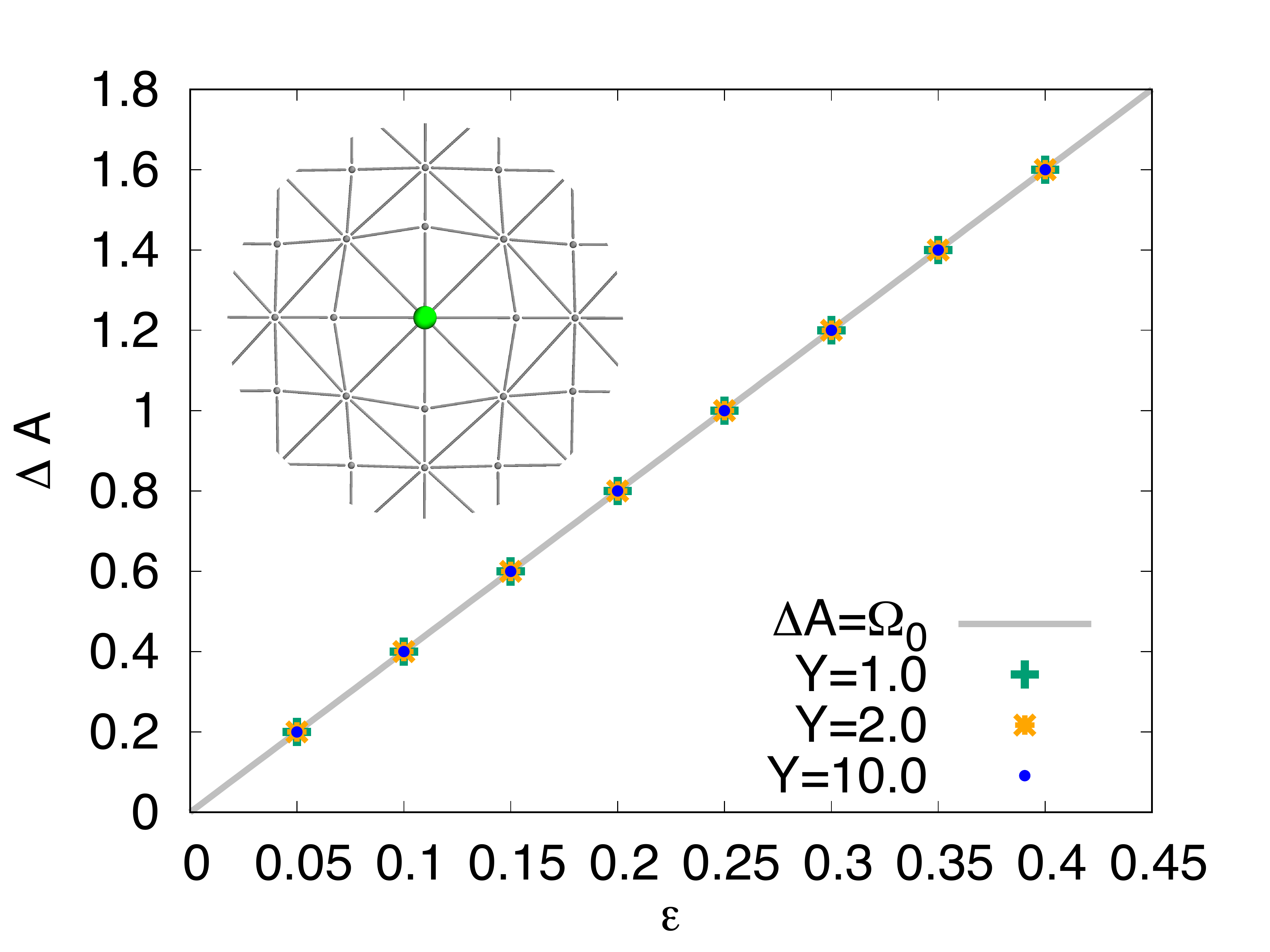}
\end{center}
\caption{{\label{2dsq} Comparison between theory [Eq. (\ref{sqomega0})] and simulation for the macroscopic increase in area due to an impurity versus the parameter $\epsilon$ for various values of the Young's modulus $Y=\frac{4}{3} k$, where $k$ is the spring constant. The gray line shows $\Delta A = 4 a_0^2\epsilon$. Data are for a square periodic cell with ``radius" (half of a side length) $R=30 a_0$. Inset: Local lattice deformations near an impurity site obtained by energy minimization in our discrete square lattice model. Compare to Fig. \ref{extarea}.
}}
\end{figure}

\new{\subsection{Discrete model design: square host lattice}}
To study this question, we design a discrete model of a 2D isotropic solid whose underlying lattice has a local square symmetry. This model allows us to have a square lattice of dilational impurities, whose postbuckling ground state we expect to be an unfrustrated checkerboard configuration. We begin with a simple square lattice, and add diagonal bonds between a subset of next-nearest-neighbor pairs to remove floppy modes. Placing bonds between all next-nearest neighbors is an appealing option, because it approximates a pair potential with a second minimum at $\sqrt{2} a_0$, thus favoring a square lattice. Unfortunately, with \textit{two} diagonal bonds in each square unit cell, the definition of the normals used to calculate bending energy is somewhat complicated. We instead make a simpler choice, shown in the inset of Fig. \ref{2dsq}, which results in uniquely defined normals, but two distinct types of sites on the lattice. Half of the sites have four bonds, all of which connect to nearest neighbors, and the other half have eight bonds, half of which connect to nearest neighbors. We place impurities only on sites that have eight bonds in order to better approximate isotropic dilations. The relative strengths of the diagonal and nearest-neighbor springs will be adjusted to produce isotropic elastic behavior at long wavelengths, despite the local square symmetry.

To fully specify the geometry, we must provide rest lengths for the bonds, which, as before, are modeled as harmonic springs. In the absence of impurities, nearest-neighbor bonds have rest length $a_0$, and diagonal next-nearest-neighbor bonds have rest length $\sqrt{2} a_0$. For a dilational impurity, we would like the $\gamma \to \infty$ ground state to be an isolated pyramid with zero stretching energy, as we found for the triangular lattice. With this aim in mind, on each impurity site we extend the nearest-neighbor bonds to have rest length $a_0(1+\epsilon)$ and the next-nearest-neighbor bonds to have rest length $a_0(\sqrt{2+2 \epsilon+ \epsilon^2})$. We can then calculate the excess area of the planar state $\Omega_0$ as before to be
\begin{equation}
\Omega_0 = \Delta A = 4 (a_0^2(1+\epsilon)-a_0^2) = 4 a_0^2 \epsilon.
\label{sqomega0}
\end{equation}
Figure \ref{2dsq} shows the linear relationship between $\Delta A$ and $\epsilon$ with the predicted slope. Unlike for the triangular lattice, we do not see any deviation at high $\epsilon$, presumably because there are no higher order corrections to $\Omega_0$ in this case. As for the triangular lattice, the magnitudes of the elastic constants drop out. 

Having specified our lattice structure, we now follow  standard methods to find the spring constants for the discrete model that approximate an isotropic solid in the continuum limit \cite{seung, landau}. While in principle the spring constants for nearest and next-nearest neighbors could differ, we find that they are the same for this model, and equal to 
\begin{equation}
k= \frac{3}{4} Y,
\end{equation}
where $Y$ is again the macroscopic 2D Young's modulus. We also find a 2D Poisson ratio $\nu=1/3$, just as we had for the triangular lattice. 

\begin{figure}
\begin{center}
\includegraphics[width=\columnwidth]{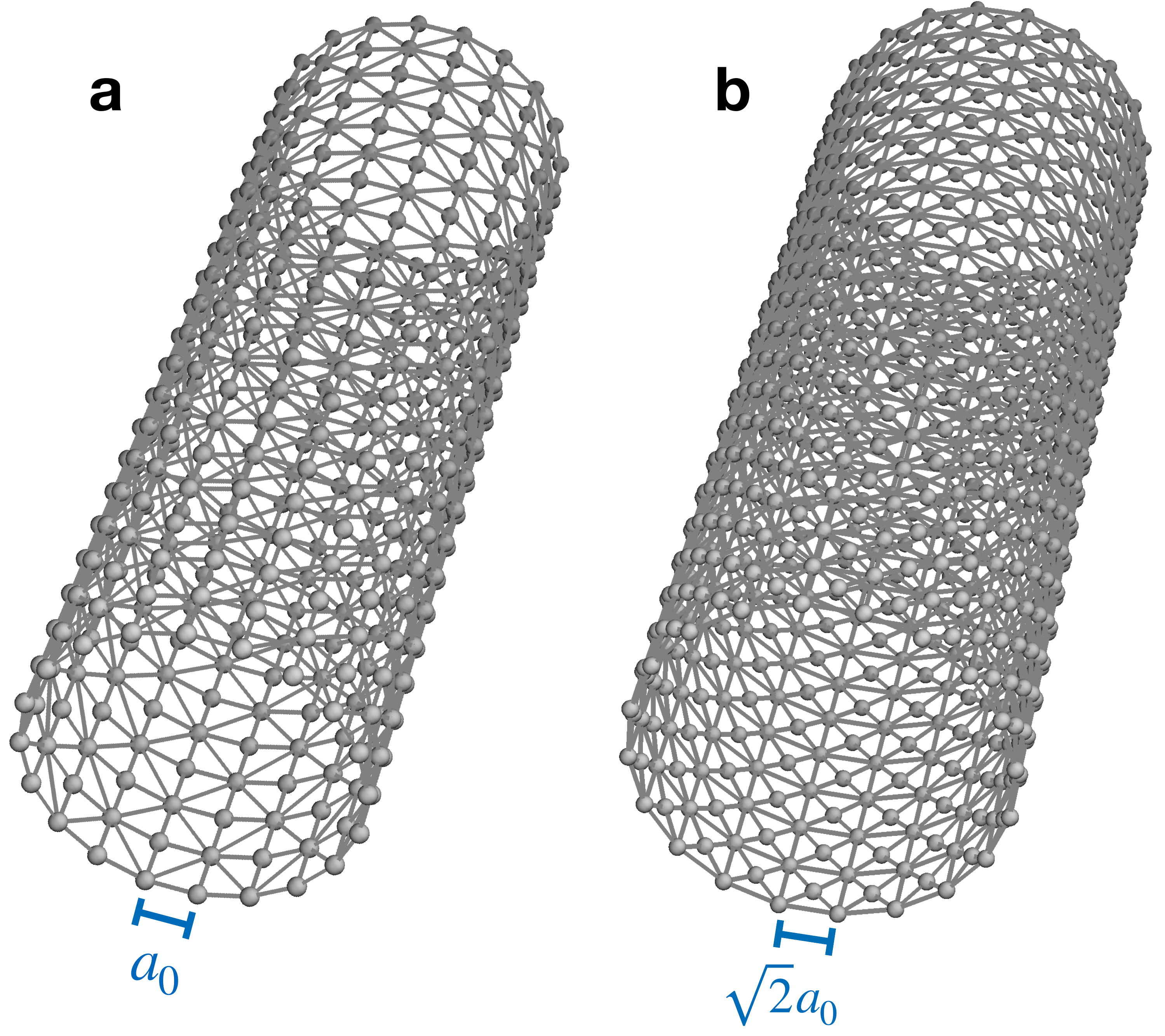}
\end{center}
\caption{{\label{cylinder} Square lattices (without dilations) rolled into cylinders such that (a) short bonds of length $a_0$ are along the axis and (b) diagonal bonds of length $\sqrt{2} a_0$ are along the axis.  
}}
\end{figure}

To find the correspondence between the discrete bending rigidity $\tilde{\kappa}$ and the continuum limit parameter $\kappa$, we roll the modified square lattice into a cylinder along a vertical axis such that faces only bend perpendicular to hinges formed by nearest-neighbor bonds of length $a_0$ [Fig. \ref{cylinder}(a)]. This configuration allows us, both analytically and numerically, to show
\begin{equation}
\tilde{\kappa}=\kappa,
\end{equation}
where $\kappa$ is the continuum value of the bending rigidity.
In principle, we could have a different value of the bending rigidity associated with bending perpendicular to bond hinges of length $\sqrt{2} a_0$. We test for this possibility by rolling a section of the lattice into a cylinder such that diagonals are along the cylinder axis, with bending now occurring across both long and short bonds [Fig. \ref{cylinder}(b)]. This construction allows us to compare the measured bending energy of large cylinders to the expected bending energy from continuum theory, and confirm that using only one value of $\tilde{\kappa}=\kappa$ gives the right result. By explicitly calculating the normals, this can also be shown analytically. Calculations and data for the bending energy as a function of $R$ for both cylinder constructions are shown in Appendix B and Fig. \ref{isobend} below. 

We can now define a dimensionless dilation F{\"o}ppl-von K{\'a}rm{\'a}n number $\gamma$ as for the triangular lattice model, differing only in the form of $\Omega_0$:
\begin{equation}
\gamma\equiv \frac{Y \Omega_0}{\kappa} = \frac{4 (4 a_0^2 \epsilon) k}{3 \tilde{\kappa}}.
\end{equation}

\begin{figure}
\begin{center}
\includegraphics[width=\columnwidth]{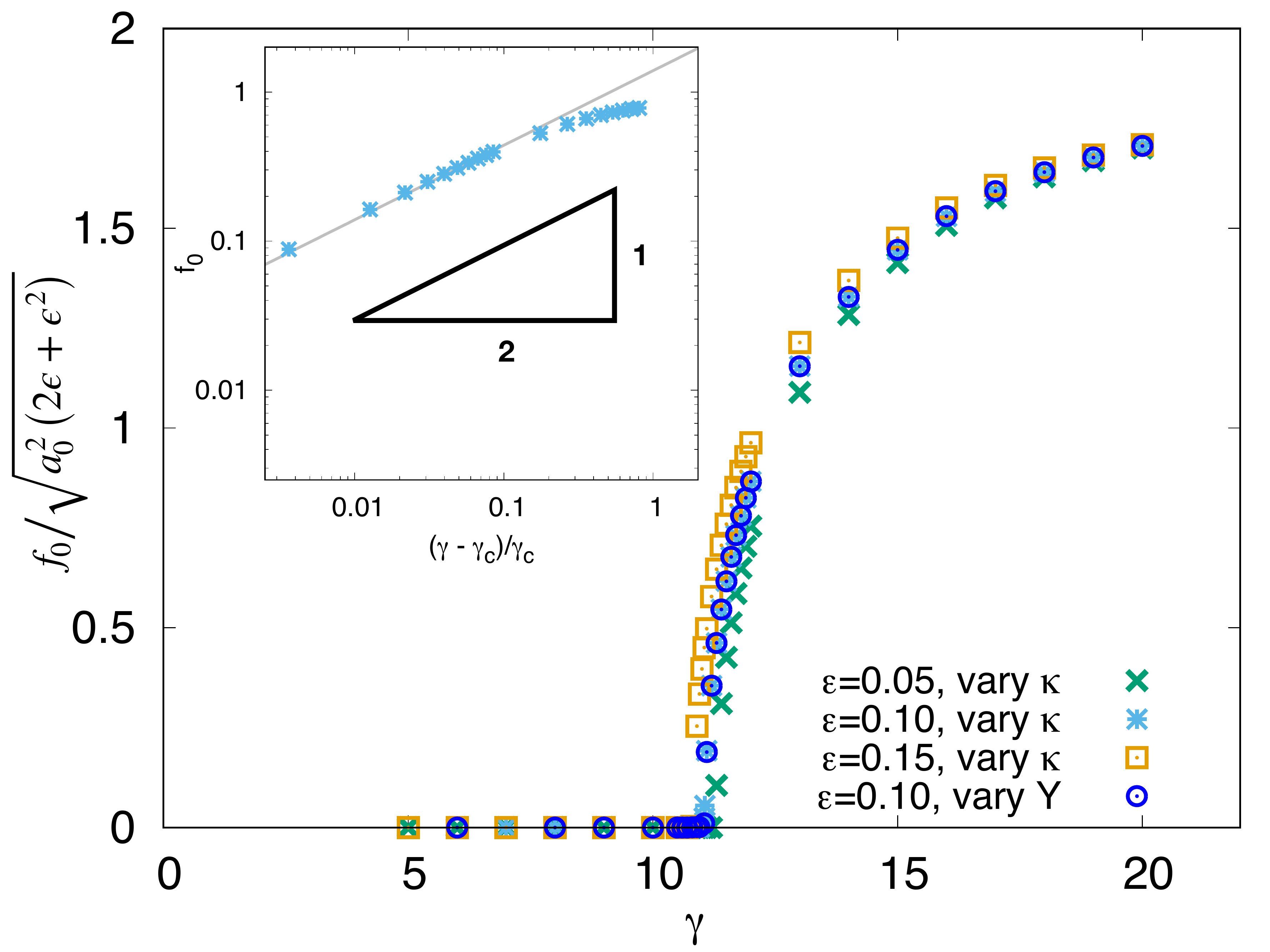}
\end{center}
\caption{{\label{scalingsq}  Height of a single impurity, rescaled by the height in the prismatic limit as a function of $\gamma$ for $R=15 a_0$. Close to the transition, the height scales as $\sqrt{\frac{\gamma}{\gamma_c}-1}$.  Inset: Data on a log-log scale as a function of $(\gamma-\gamma_c)/\gamma_c$ with $\epsilon=0.1$, and $\gamma$ changed by varying $\kappa$. Compare to Fig. \ref{collapse} and Fig. \ref{scaling}.
}}
\end{figure}

\new{\subsection{Square lattice results}}
As before, we find a buckling transition for an isolated dilation with increasing $\gamma$, as shown in Fig. \ref{scalingsq}, and impurity height scaling close to the transition that again goes as $\sqrt{(\gamma-\gamma_c)/\gamma_c}$ (inset of Fig. \ref{scalingsq}). 

\begin{figure}
\begin{center}
\includegraphics[width=\columnwidth]{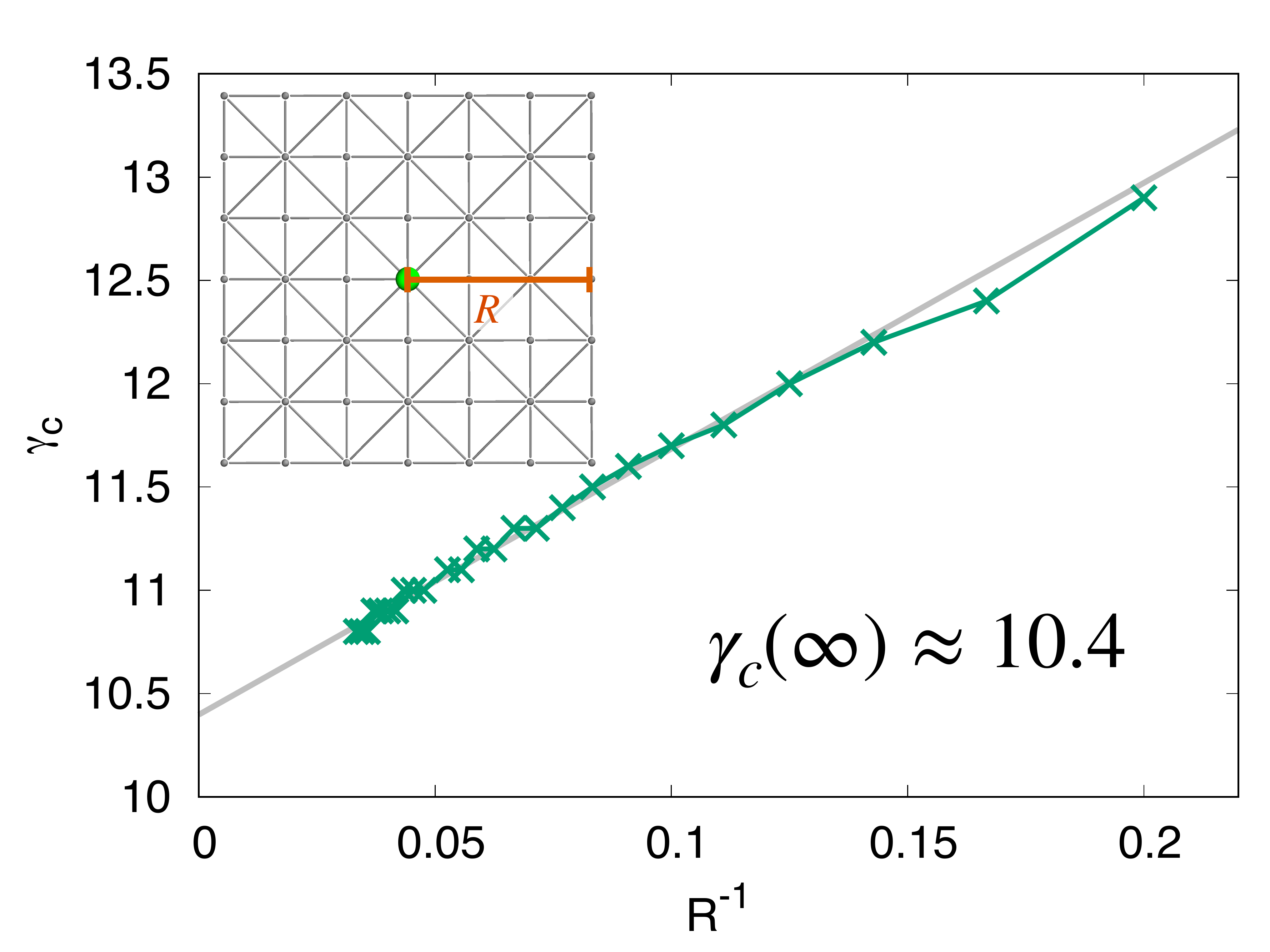}
\end{center}
\caption{{\label{vlsq} Variation of $\gamma_c$ for the buckling of a single dilation and its periodic images with $R=n a_0$. When plotted as a function of $R^{-1}$ (units of $a_0^{-1}$), a linear extrapolation to the infinite system size limit gives $\gamma_c \approx 10.4$. Data are shown for $n$ between $5$ and $30$, $\epsilon=0.05$, and $\gamma$ is changed by varying $\kappa$. Compare to Fig. \ref{vl}.
Inset: Periodic unit cell of the defect superlattice for $R=3a_0$.
}}
\end{figure}

\begin{figure}
\begin{center}
\includegraphics[width=\columnwidth]{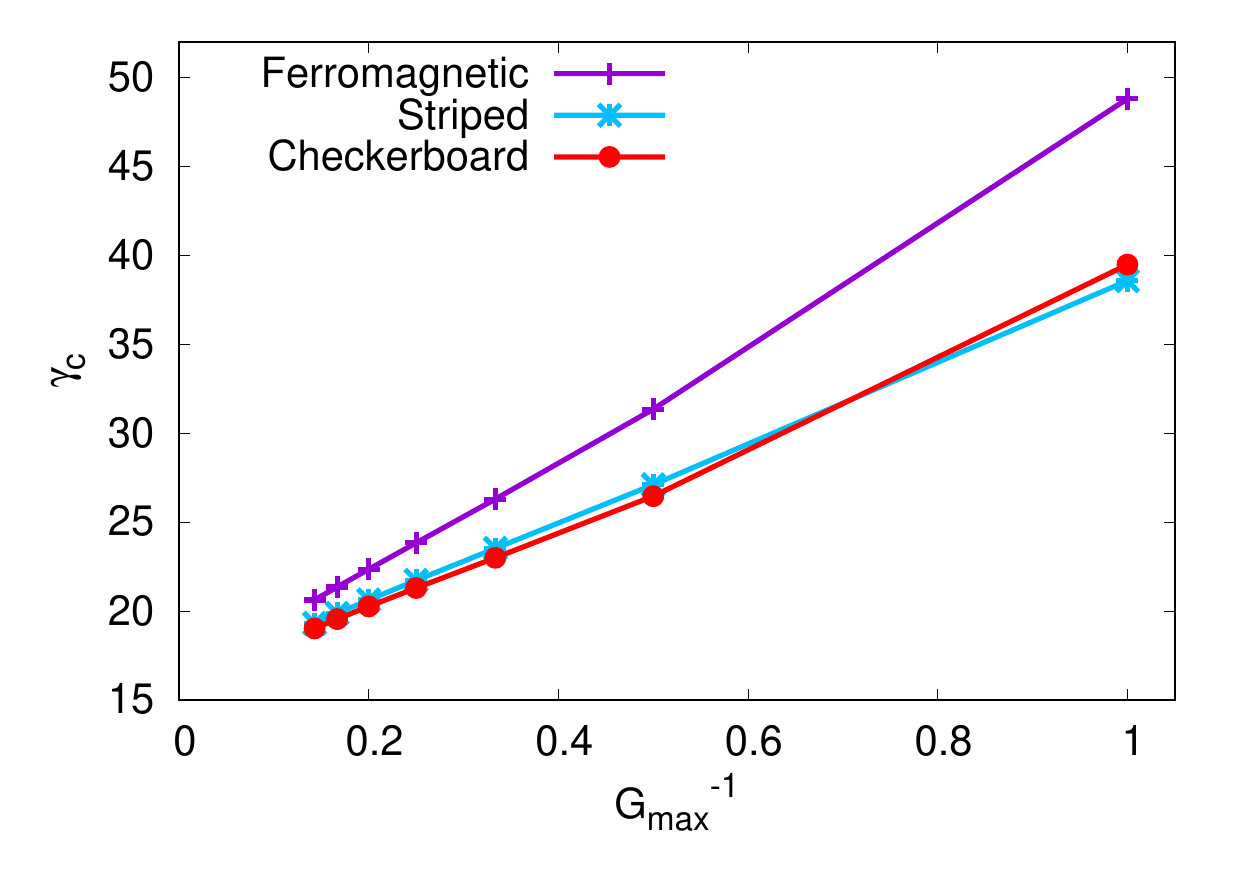}
\end{center}
\caption{{\label{comparesq} Variation of $\gamma_c$ with $G_{\text{max}}=n g_0$ for the ferromagnetic state, striped state, and checkerboard state. Results are shown for $n$ between 1 and 7, with $G_{\text{max}}$ measured in units of $g_0$. Compare to Fig. \ref{compare}.
}}
\end{figure}

We can again systematically increase the size of the discrete system and the number of modes in the Fourier space theory to estimate the buckling threshold in the infinite system size limit. For the square lattice, the periodic image impurities appear along the $(0,n)$ direction. We find rough agreement between the measured values of $\gamma_c$ shown in Fig. \ref{vlsq} [$1/R$ extrapolation $\gamma_c(\infty)\approx 10.4$] and the predicted values shown by the purple line in Fig. \ref{comparesq} below [$1/G_{max}$ extrapolation $\gamma_c(\infty)\approx 16.4$]. As in the triangular lattice case, the simulation value is lower than the continuum elastic theory estimate.  Note that we quote lattice size in terms of $R$, which we define to be half of the length of a side of the square domain defining our periodic boundary conditions (inset of Fig. \ref{vlsq}), for the purpose of easier comparison with the hexagonal domain used for the triangular lattice study. 

Finally, we study periodic arrays of interacting dilations inserted into our square host lattice. By analogy with an Ising model on a square lattice with nearest and next-nearest-neighbor interactions, we focus on only three candidate ground states \cite{kassan}. The checkerboard state, a N\'{e}el configuration shown in Fig. \ref{checker}, will be the ground state if (as we expect for our system) there is strong nearest-neighbor antiferromagnetism. The ferromagnetic state [Fig. \ref{dEsq}(b)] would have the lowest energy if ferromagnetic nearest-neighbor interactions were to dominate, and the striped state [Fig. \ref{dEsq}(a)], would have the lowest energy if next-nearest-neighbor antiferromagnetism were to dominate. Other ground states are possible with strong third-nearest-neighbor interactions \cite{kassan}, but these do not appear to be realized for our model. 

\begin{figure}
\begin{center}
\includegraphics[scale=0.28]{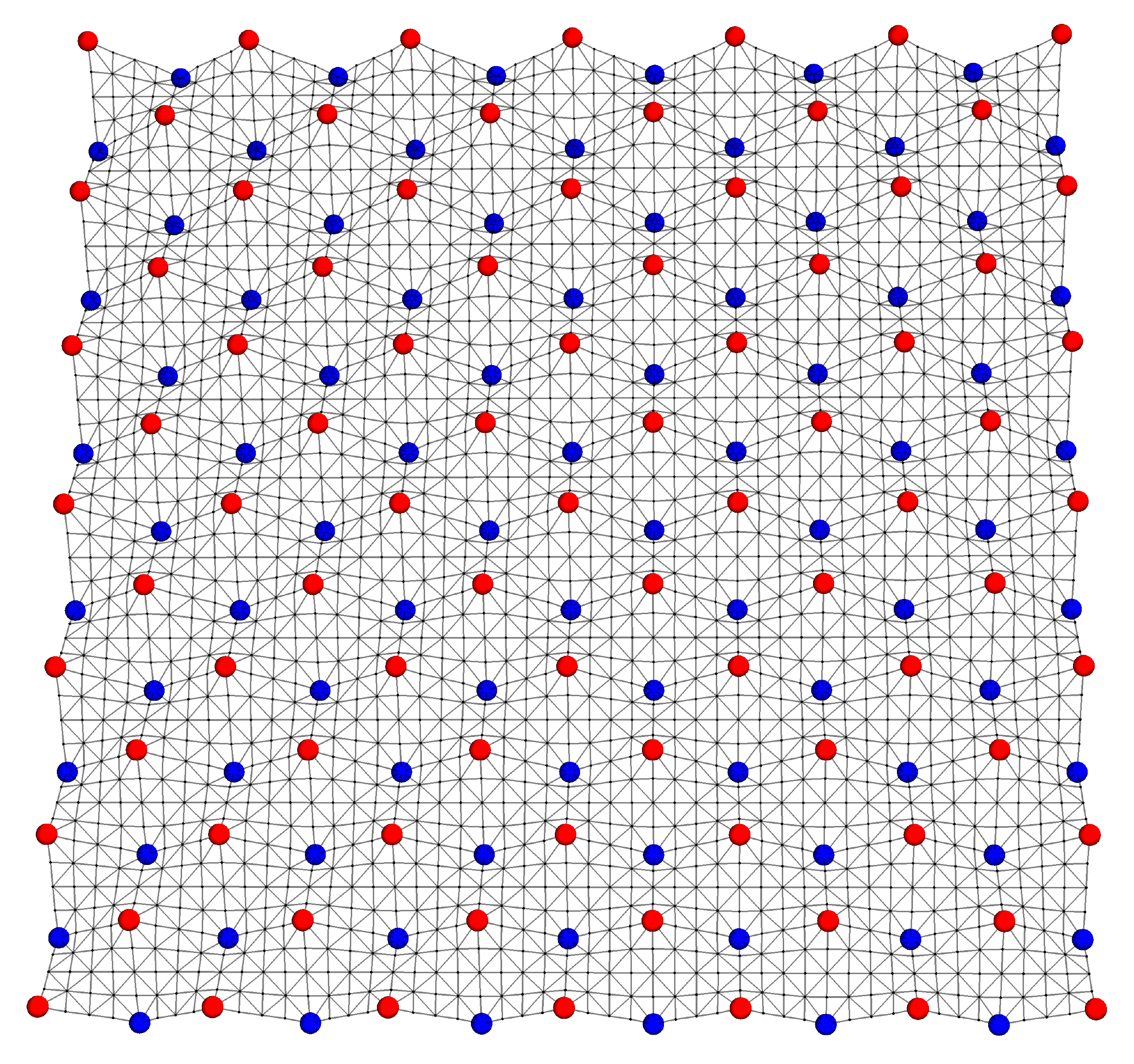}
\end{center}
\caption{{\label{checker} The checkerboard state for a $(0,4)$ array in a square periodic domain of size $R=24 a_0$ at $\gamma=26.7>\gamma_c$, viewed in perspective with the vertical displacements magnified by a factor of 5 for clarity. This configuration is our conjectured ground state for the square lattice. Impurities that have buckled up are shown in red, and impurities that have buckled down are shown in blue. Compare to Fig. \ref{zigzag}.
}}
\end{figure}

We restrict our attention to $(0,n)$ arrays, i.e., dilational impurities separated along a row of nearest-neighbor bonds. These arrays seem to be robust to microscopic lattice details, and allow us to make contact with our results for a single impurity with periodic boundary conditions, just as $(n,n)$ arrays did for the triangular lattice case.

\begin{figure}
\begin{center}
\includegraphics[width=\columnwidth]{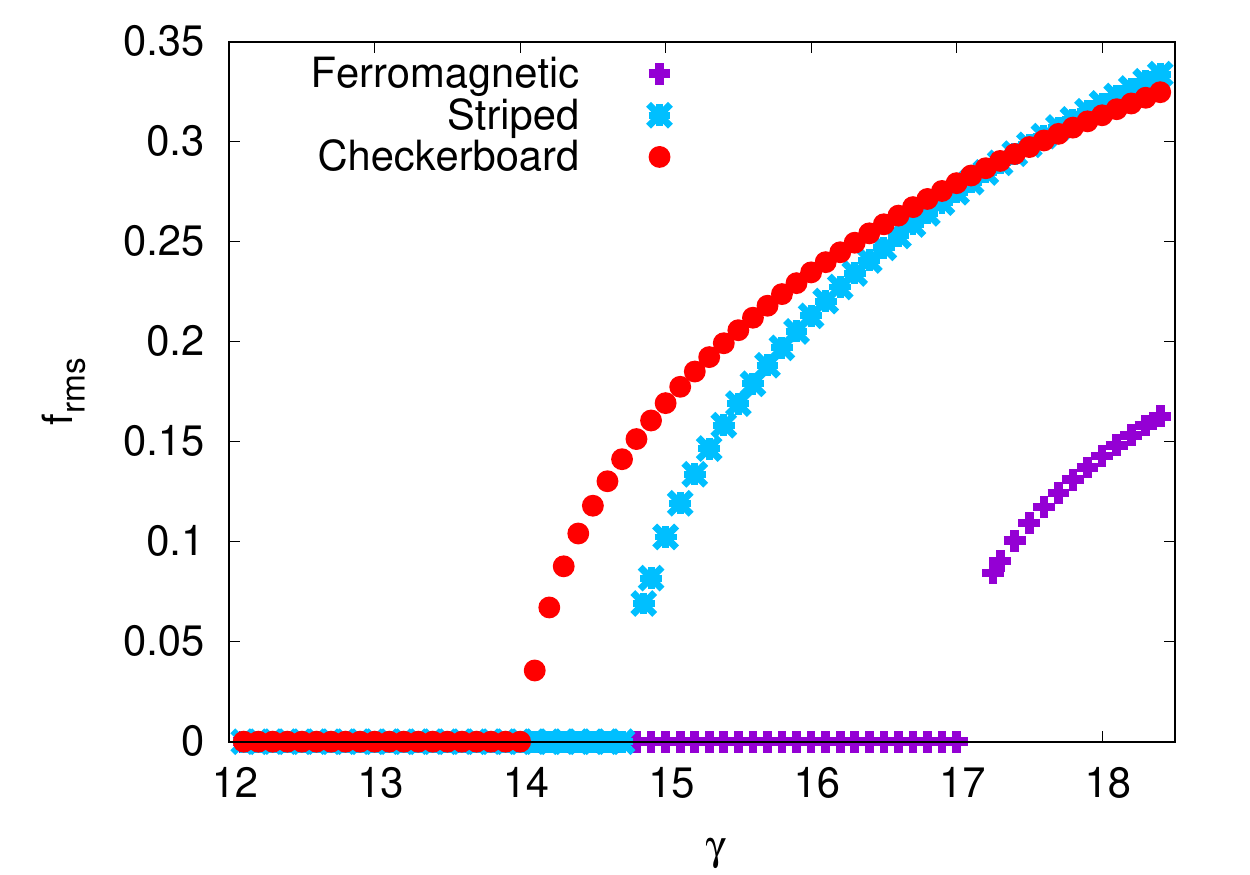}
\end{center}
\caption{{\label{bucklesq} Root-mean-square height measured at impurity sites versus $\gamma$. The checkerboard state buckles first, followed by the striped state, and then the ferromagnetic state. Data are for $(0,4)$ arrays with $R=96a_0$, $\epsilon=0.1$, and $\gamma$ is changed by varying $\kappa$. Compare to Fig. \ref{buckle}. 
}}
\end{figure}

As anticipated, the checkerboard state appears to be the postbuckling ground state for all values of $\gamma$ by a more substantial margin than in the triangular case (Fig. \ref{dEsq}), and is the first configuration to buckle (Fig. \ref{bucklesq}). Note that there are gaps in the data in Fig. \ref{bucklesq} for the striped and ferromagnetic states close to the transition. In these regions, the striped or ferromagnetic states lose their metastability and ``spins" are able to flip during energy minimization and create domains of the more stable checkerboard phase.

We can also estimate $\gamma_c$ for these three states using the Fourier space continuum elastic theory employed for a triangular host lattice. As shown in Fig. \ref{comparesq}, the checkerboard state has the lowest buckling threshold when more than eight Fourier modes are included in the calculation, as well as when we extrapolate to infinite system size. Upon extrapolating to $G_{\text{max}}^{-1}=0$ using a linear fit that neglects the two points with the fewest Fourier modes, we find an estimate of $\gamma_c(\infty)=16.4$ for the ferromagnetic state, $\gamma_c(\infty)=16.2$ for the striped state, and $\gamma_c(\infty)=16.1$ for the checkerboard state.

\begin{figure}
\begin{center}
\includegraphics[width=\columnwidth]{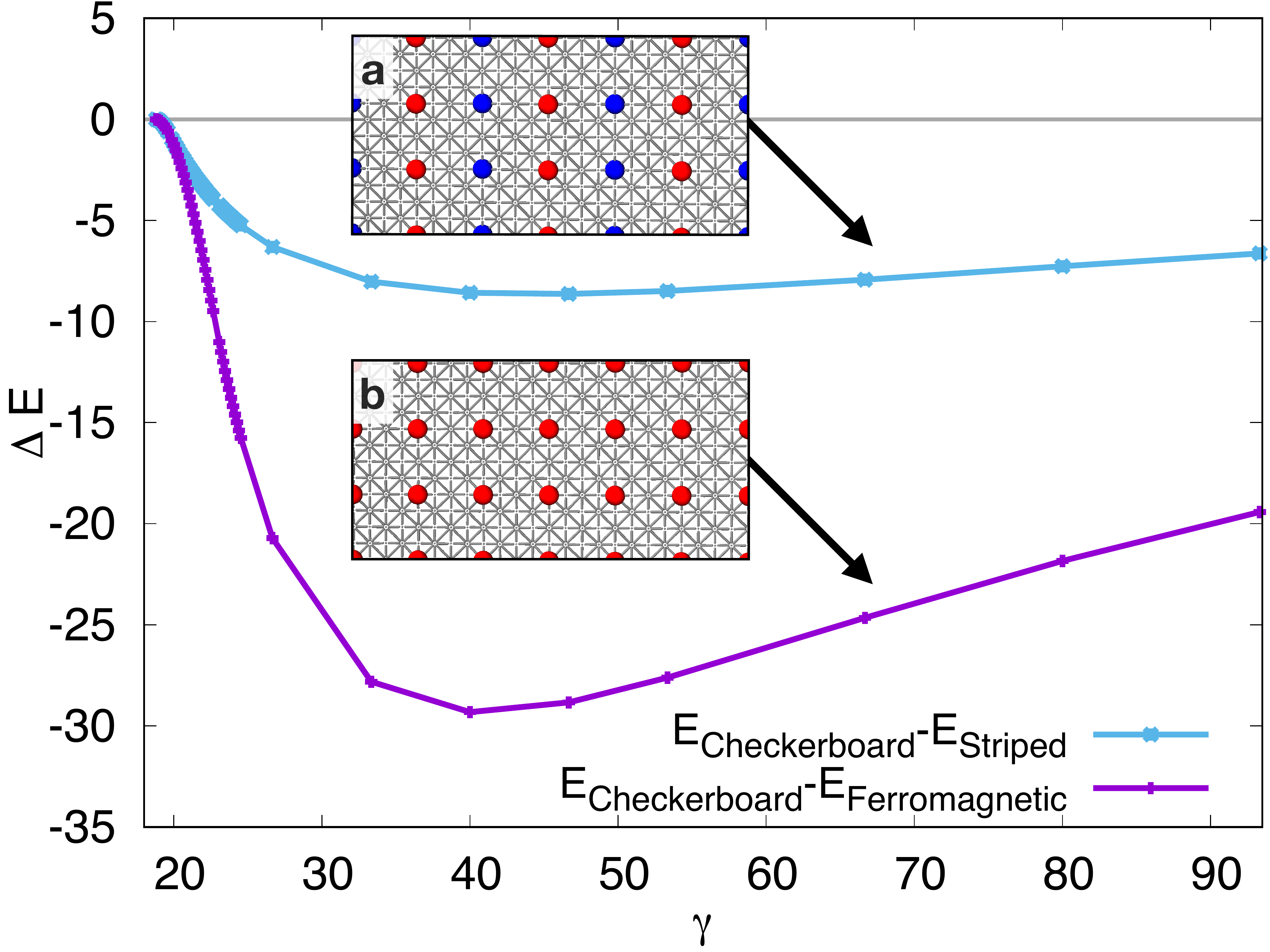}
\end{center}
\caption{{\label{dEsq} Difference in energy between two metastable states, the striped state [(a), stripes run vertically] and the ferromagnetic state (b), and our conjectured checkerboard ground state pictured in Fig. \ref{checker} for $\gamma$ larger than $\gamma_c$ of the checkerboard state. All data are for $(0,4)$ arrays with $R=96a_0$, $\epsilon=0.1$, and $\gamma$ is changed by varying $\kappa$. Compare to Fig. \ref{dE}, and note the difference in scale on the $y$ axis. }}
\end{figure}

\section{DISCUSSION}
\label{discussion}

The dilational impurity arrays studied here provide a simple arena for exploring shape memory, instability, metastable states, and Ising-like phase transitions for 2D surfaces embedded in three dimensions. Tools from continuum elastic theory allow us to predict the approximate location and nature of the buckling transitions as a function of the dilation F{\"o}ppl-von K{\'a}rm{\'a}n number $\gamma= \frac{Y \Omega_0}{\kappa}$. Furthermore, we have conjectured ground states for $(n,n)$ arrays embedded in triangular host lattices and $(0,n)$ arrays with square host lattices, and found that these conjectures are consistent with simulations, calculations, and an Ising model analogy.

The Ising analogy provided us with a set of states to test, and in our analysis we assumed that the true ground state was contained in this set. While this approach has an appealing physical motivation, it would also be of interest to design a numerical experiment allowing the system itself to select the ground state. We could, for example, start a triangular lattice with dilations at a high temperature, slowly cool it down, and hope to observe domains of the zigzag state forming. However, such simulations are technically challenging because of the large system size required, the many metastable states with energies close to the ground state, especially for the triangular lattice, and the energy barrier for flipping an impurity and nearby affected sites from up to down. Although beyond the scope of this work, these difficulties could certainly be overcome, and we feel that this avenue is worth pursuing.

An especially intriguing aspect of finite temperature simulations of a regular lattice of dilations embedded in one of the ``tethered surfaces" studied here is the interplay between crumpling transitions and the Ising-like orderings of buckled dilations. Even in the absence of an array of buckled impurities, triangulated surfaces in a low-temperature flat phase can undergo a transition to an entropy-dominated crumpled state above a critical temperature $T_c$ \cite{nelsonbook}. Recent results suggest that this crumpling temperature (in the absence of distant self-avoidance) can be lowered considerably by inserting a regular array of holes in a simple model of free-standing graphene \cite{yllanes}. If these holes are replaced by the regular array of dilations studied here, the resulting membranes have additional internal degrees of freedom. There can now be a finite temperature buckling transition at a temperature $T_b$, as well as a transition to a low-temperature phase where these puckers then order into a zigzag or checkerboard state via an Ising model phase transition at temperature $T_I$. It would be intriguing to study such phase transitions in this system, which resembles a highly compressible Ising model: The host lattice itself may rise up and crumple for entropic reasons! Self-avoidance of the host polymer sheet might well play a role under some circumstances \cite{paczuski}, as would the relative ordering of important temperatures such as $T_c$, $T_b$, and $T_I$.

Although we have focused here on dilations, which locally add extra area to the lattice, one could also study defect arrays that remove area, such as a lattice of vacancies. In 2D flat space, removing a single particle from a triangular array typically produces a ``crushed vacancy,” whose elastic field has a dipolar character \cite{jain}. Little is known about what happens to interacting arrays of crushed vacancy dipoles, especially when allowed to relax into the third dimension.   

It would also be of interest to study more systematically other $(n,m)$ tessellations of dilations in the host lattice, particularly the chiral versions with $n \neq 0$, $m \neq 0$, and $n \neq m$, which play an important role in the capsids of viruses \cite{casparklug} and in various phyllotaxis problems (see Ref. \cite{beller} and references therein). 

Finally, while we studied systems with periodic boundaries to better approximate an infinite, approximately planar material, experimental realizations of this system and related systems will likely have free or clamped boundaries. Boundary effects should therefore be systematically studied, as they can profoundly change the behavior in small systems, as illustrated by Fig. \ref{cappbc}(a).

\begin{acknowledgments}
We thank Michael Moshe and Paul Hanakata for useful discussions. This work was supported by the National Science Foundation, through grant DMR-1608501 and via the Harvard Materials Science Research and Engineering Center through grant DMR-2011754.
\end{acknowledgments}

\appendix
\section*{Appendix A: Results for dilation buckling with an exponential profile}
\label{exponential}
We present here a back-of-the-envelope calculation to motivate the existence of a buckling threshold like that in Eq. (\ref{gammaintro}), using, however, an exponential height profile that better mimics details of our discrete model than a Gaussian [see Figs. \ref{collapse}(a) and \ref{collapse}(b)]. 
Upon assuming a buckled state height profile of the form
\begin{equation}
f(r)= H_0 e^{-r/\sigma},
\end{equation}
the bending energy is
\begin{equation}
E_b = \frac{\kappa H_0^2}{2} \int d^2r \frac{e^{- 2 r/\sigma} ( \sigma - r)^2}{\sigma^4 r^2}.
\end{equation}
If we integrate this from $r=\delta$, a microscopic cutoff, to $r=\infty$ , the result can be expressed in terms of the exponential integral function $\text{Ei}(z)= -\int_{-z}^{\infty} e^{-t}/t dt$,
\begin{equation}
E_b=\pi \kappa H_0^2\frac{ e^{-2\delta/\sigma} (2 \delta - 3 \sigma) - 4 \sigma \text{Ei}( -2 \delta/\sigma)}{4 \sigma^3}.
\end{equation}
Upon letting $x= 2 \delta/\sigma$, this simplifies to
\begin{equation}
E_b=\frac{\pi \kappa H_0^2}{4 \sigma^2} ( e^{-x} (x - 3) - 4 \text{Ei}( -x)).
\end{equation}
\new{We require that both $\delta$ and $\sigma$ are of order a few lattice constants $a_0$, so that there is negligible stretching energy outside the inclusion core. We therefore assume that the term in parentheses provides a multiplicative constant. }

As before, we relate $\Omega_0$ to $H_0$ by considering the extra surface area generated by the dilation in the Monge representation (now integrating from $r=0$ to $r=\infty$). This gives
\begin{equation}
\Omega_0 \approx \frac{\pi H_0^2}{4}.
\end{equation}
If we again ask when the bending energy of a buckled dilation is lower than the stretching energy [Eq. (\ref{ur})], we once again find that it occurs above a critical value of the dimensionless F{\"o}ppl-von K{\'a}rm{\'a}n number $\gamma$, constructed with the impurity size $\Omega_0$,
\begin{equation}
\gamma_c \equiv \frac{Y \Omega_0}{\kappa} \sim \frac{\delta^2}{\sigma^2},
\end{equation}
which agrees with Eq. (\ref{gammaintro}). We have assumed here that the microscopic bending and stretching cutoffs are both $\delta$.

We can also superimpose two exponential height profiles and calculate the difference in bending energy between the ferromagnetic and antiferromagnetic configurations, as we did for Gaussian profiles in Eq. (\ref{gaussbend}). The superimposed exponential height profile [compare with Eq. (\ref{2gauss})] is now
\begin{equation}
f(x,y)=  H_0\left(e^{-\sqrt{(x - d)^2 + y^2 }/\sigma}\pm e^{-\sqrt{(x+ d)^2 + y^2 }/\sigma} \right).
\end{equation}
We find the difference in bending energy between the anti-aligned and aligned configurations [$E_{+-}-E_{++}$, as in Eq. (\ref{gaussbend})] by numerically integrating $(\nabla^2 f)^2$ over a large region that excludes an area of radius $\delta$ around $x=\pm d, y=0$. As in the case of superimposed Gaussians, we find short-range ferromagnetism and long-range antiferromagnetism if we only consider bending energy. 

\section*{Appendix B: Bending energy of square lattice cylinders}
\label{sqbending}
Following standard techniques \cite{seung, landau}, we expect the continuum limit of the bending energy for a cylinder with bending rigidity $\kappa$ to be
\begin{equation}
E_b= \frac{\kappa \pi L}{R},
\label{contcylinder}
\end{equation}
where $R$ and $L$ are the cylinder radius and length, respectively. For the case that we simulate, in which the cylinder is constructed by rolling up a square domain so that $R= L/2 \pi$, this reduces to 
\begin{equation}
E_b = 2 \kappa \pi^2.
\end{equation}

Using the discrete formulation, not yet assuming that the bending rigidity associated with hinges of length $\sqrt{2}a_0$, $\tilde{\kappa}^\prime$, is equal to the bending rigidity associated with hinges of length $a_0$, $\tilde{\kappa}$, the bending energy can be written
\begin{equation}
E_b=\tilde{\kappa} \sum_{<\alpha \beta>} (1- \mathbf{n}_\alpha \cdot \mathbf{n}_\beta)+ \tilde{\kappa}^\prime \sum_{<\gamma \delta>} (1- \mathbf{n}_\gamma \cdot \mathbf{n}_\delta),
\end{equation}
where $\alpha$ and $\beta$ index neighboring triangular faces sharing an edge of rest length $a_0$, and $\gamma$ and $\delta$ index faces sharing an edge of rest length $\sqrt{2}a_0$. We now show that $\kappa=\tilde{\kappa}=\tilde{\kappa}^\prime$ in the continuum limit. 

\subsection*{Cylinder in Fig. \ref{cylinder}(a)}
In the configuration shown in Fig. \ref{cylinder}(a), all bending occurs across hinges formed by short bonds (rest length $a_0$). If we look down the axis of the cylinder, we see a circle with radius $R$ that is discretized by $N$ points, such that $N a_0=2 \pi R$. Neighboring normals are rotated by an amount 
\begin{equation}
\Delta \theta= \frac{2 \pi}{N} = \frac{a_0}{R}.
\end{equation}
Counting up all contributing pairs of normals, the total bending energy in the limit of large $N$ is
\begin{equation}
E_b= \tilde{\kappa} \frac{N L}{a_0} (1- \cos (\Delta \theta))\approx \frac{\tilde{\kappa} \pi L}{R}.
\end{equation}

This is equal to the continuum result (\ref{contcylinder}) when $\tilde{\kappa}= \kappa$.

\subsection*{Cylinder in Fig. \ref{cylinder}(b)}
We now consider the configuration shown in Fig. \ref{cylinder}(b). There are two types of sites in our square lattice: those with four short bonds and those with eight bonds, four short and four long. If the sites with only four short bonds are omitted, we form a simple square lattice with lattice constant $\sqrt{2} a_0$. If we were to roll the simple square lattice into a cylinder and look down its axis, we would see a circle discretized by $N= \frac{2 \pi R}{\sqrt{2} a_0}$ points. However, when we also include the sites with four short bonds, we find a circle discretized by twice as many points, and $N= \frac{4 \pi R}{\sqrt{2}a_0}$. In other words, each unit cell of the simple square lattice has a node with four short bonds at its center, offset radially from its four nearest neighbors. This locally forms a pyramid in each unit cell, with the pyramid height going to zero as $R \to \infty$. 

There are three types of contributions to the bending energy. The first is similar to that considered above for the cylinder in Fig. \ref{cylinder}(a): bending occurs perpendicular to long hinges due to the rotation of the normals by an amount $\Delta \theta= \frac{2 \pi}{N}$. Counting up all the relevant pairs of normals, this type of bending contributes $\tilde{\kappa}^\prime \frac{\pi L}{4 R}$ to the total bending energy. 

The second type of bending energy comes from the bending across short hinges in the pyramid unit cells. In coordinates such that $\hat{z}$ is the radial displacement, the four points at the base of the cylinder are 
\begin{equation}
\*r_i= \left( \pm R \sin \left(\Delta \theta\right), \pm \frac{\sqrt{2}}{2}a_0,0 \right),
\end{equation}
and the point at the peak is
\begin{equation}
\*r_0= \left( 0, 0, R\left(1- \cos\left( \Delta \theta \right) \right)\right).
\end{equation}
Note that if we take the continuum limit by sending $N \to \infty$ with $Na_0$ fixed, so that $R$ remains constant, we find that the $\{\*r_i\}$ points are at the corners of a square of side length $\sqrt{2} a_0$, as expected. 

We then can explicitly compute the normals for the four faces of the pyramid and calculate the bending energy for each pair of faces that share an edge. We expand the result in the large $N$ limit, keeping only the lowest order terms in $1/N$. Each of the $\frac{2 \pi R L}{2 a_0^2}$ pyramids contribute identically to the bending energy. The total energy from this type of bending is $\tilde{\kappa} \frac{\pi L}{2 R}$. 

The third type of bending energy comes from bending across long hinges. If we consider a line of pyramids along the axis of the cylinder, the long hinges joining the pyramids together also contribute bending energy. This term, to lowest order in $1/N$ gives a total contribution of $\tilde{\kappa}^\prime \frac{\pi L}{4 R}$. 

Summing up the three contributions, we find
\begin{equation}
E_b = \tilde{\kappa}^\prime \frac{\pi L}{2 R}+\tilde{\kappa} \frac{\pi L}{2 R}.
\end{equation}
We already know that $\tilde{\kappa}= \kappa$ from the cylinder in Fig. \ref{cylinder}(a), so we conclude that $\tilde{\kappa}^\prime= \kappa$ as well.

\begin{figure}
\begin{center}
\includegraphics[width=\columnwidth]{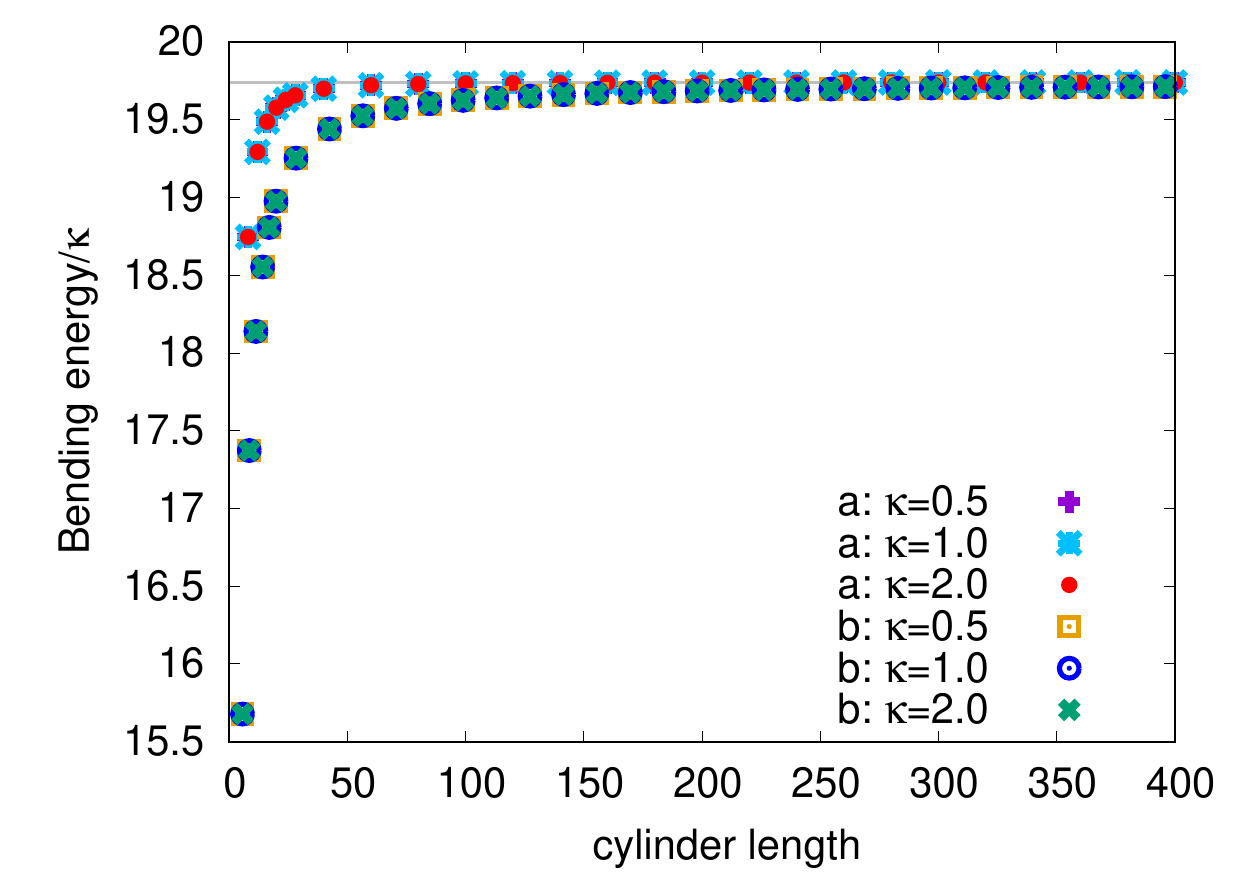}
\end{center}
\caption{{\label{isobend} As the cylinder size is increased, the bending energy approaches $E_b=2 \kappa \pi^2$ for both types of cylinders shown in Fig. \ref{cylinder}. Cylinders constructed as in Fig. \ref{cylinder}(a) are labeled ``a," and cylinders constructed as in Fig. \ref{cylinder}(b) are labeled ``b." Note that the cylinder length $L$ is always equal to the cylinder circumference $2\pi R$ in these simulations.
}}
\end{figure}

For both types of cylinders pictured in Fig. \ref{cylinder}, we also confirm numerically that the bending energy approaches the expected value for large cylinders with $\kappa=\tilde{\kappa}= \tilde{\kappa}^\prime$ in Fig. \ref{isobend}. 

\section*{Appendix C: Candidate ground states for the triangular host lattice}

\begin{figure}
\begin{center}
\includegraphics[width=\columnwidth]{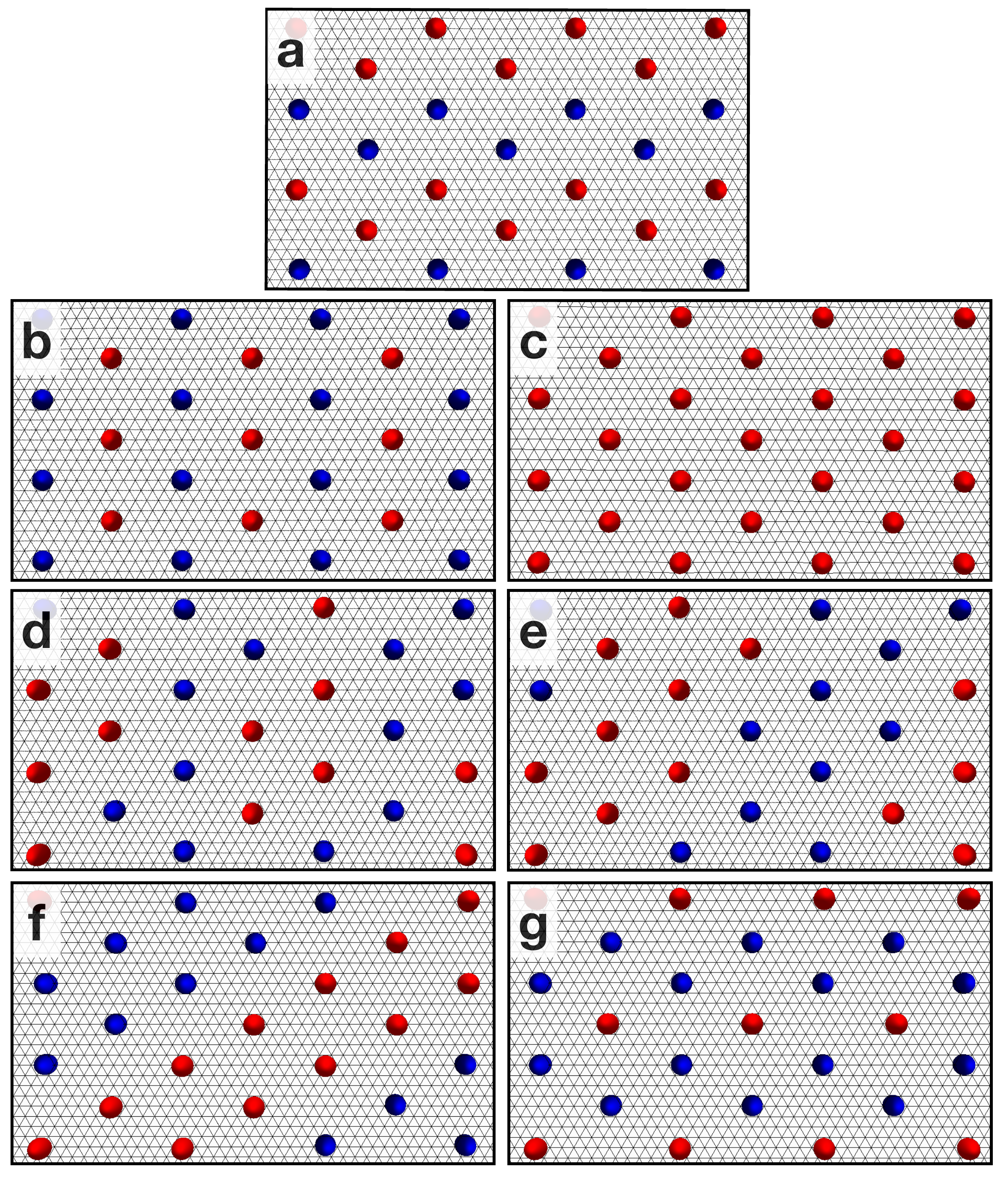}
\end{center}
\caption{{\label{isingstates} \new{Top-down views of the candidate ground states inspired by an Ising model with first, second, and third-nearest-neighbor interactions on a triangular lattice \cite{tanaka}. Impurities that have buckled up are shown in red and impurities that have buckled down are shown in blue, on a $(4,4)$ lattice. Panel (a) is our conjectured ground state for the buckled impurity array system.}
}}
\end{figure}

\new{We present the top-down views of the seven candidate ground states that we test for the triangular host lattice case in Fig. \ref{isingstates}. See \citet{tanaka} for details.}

\end{document}